\documentclass[aps,prd,preprint,nofootinbib]{revtex4}
\usepackage{amsfonts,mathrsfs,graphicx,amsmath,amssymb,slashed,float}
\usepackage{caption}
\usepackage{bm}

\newcommand{\beq}{\begin{equation}} \newcommand{\eeq}{\end{equation}}
\newcommand{\Mcal}{\mathcal{M}} \allowdisplaybreaks[1]
\begin{document}

\title{Finite-Width Dissolution of Radial Spectroscopy in\\
Single-Top Mesonic Correlations}

\author{Bing-Dong Wan$^{1,2}$}
\email{wanbd@lnnu.edu.cn}
\author{Shuo Yang$^{1,2}$}
\email{shuoyang@lnnu.edu.cn}
\affiliation{$^1$School of Physics and Electronic Technology, Liaoning Normal
University, Dalian 116029, China\\
$^2$Center for Theoretical and Experimental High Energy Physics,
Liaoning Normal University, Dalian 116029, China}

\begin{abstract}
Within the heavy-mass expansion, the pole width of a system containing one unstable heavy constituent inherits the constituent width up to $\mathcal O(\Lambda_{\rm kin}^2/m_Q^2)$ corrections, while radial splittings remain $\mathcal O(\Lambda_{\rm rad})$. The top quark is an extreme realization of this hierarchy. We implement the complex top pole mass in an instantaneous Bethe--Salpeter framework, where a biorthogonal Hellmann--Feynman relation realizes width inheritance at the operator level and an artificial heavy-mass scan confirms the predicted $m_Q^{-2}$ suppression. The low-pole source-projected response has a single broad maximum at the physical top width in the $t\bar b$, $t\bar c$, and $t\bar u$ channels. Full width-dependent non-Hermitian re-diagonalization and a direct full-matrix resolvent evaluation confirm the progressive dissolution of the small-width radial maxima. Thus stable-top eigenvalues survive as reference poles but not as a resolvable multi-peak spectrum; they may instead leave qualitative, process-dependent $Wb\bar q$ signatures, such as a broad threshold enhancement or modified color flow.
\end{abstract}

\maketitle

\section{Introduction}\label{sec:intro}

Due to the excellent detector performance and advances in theoretical calculation, recent analyses by the ATLAS and 
CMS Collaborations, based on LHC data at $\sqrt{s}=13~\text{TeV}$, have revealed a significant excess of $t\bar t$ events 
near the kinematic threshold~\cite{CMS:2025kzt,ATLAS:2026dbe}. This excess is consistent with the production of a 
color-singlet pseudoscalar quasi-bound state, as predicted by non-relativistic QCD (NRQCD) calculations for toponium
~\cite{Beneke:2013jia, Beneke:2024sfa, Garzelli:2024uhe, Shao:2025dzw, Fuks:2021xje, Fuks:2024yjj, Aguilar-Saavedra:2024mnm}.
Following the recent observations of near-threshold enhancements consistent with color-singlet $t\bar t$ quasi-bound-state dynamics, a natural question is whether additional top-quark bound states exist. Spurred by 
these developments, extensive theoretical studies have explored potential new top-quark bound states, including top-flavored 
mesons and baryons~\cite{Zhang:2025fdp,Zhang:2025bar,Luo:2025had,Zhang:2026top,Corcella:2026tnu}.

The basic scale observation that the top-quark width exceeds the
excitation scale of top-flavoured heavy-light systems was already
recognized in early potential-model
studies~\cite{Fabiano:1997un,Bigi:1986jk}.  The present advance is not
the scale comparison itself, but a heavy-mass-expansion statement for
unstable one-heavy-constituent poles, together with an operator-level
non-Hermitian realization and a direct numerical verification of the
predicted $m_Q^{-2}$ width-inheritance correction.

Notably, the top quark is qualitatively different from all other quarks because
its weak decay width, $\Gamma_t\simeq1.42$~GeV~\cite{ParticleDataGroup:2024cfk}, is larger
than a typical nonperturbative single-top mesonic excitation scale.
The resulting lifetime is much shorter than the usual QCD hadronization
time~\cite{Bigi:1986jk}. Consequently, a calculation with a stable top
quark may define useful reference positions for single-top correlations,
but its discrete eigenvalues cannot automatically be interpreted as
conventional hadron masses.

Although a physical top quark does not survive long enough to form a
conventional asymptotic hadron, stable-top bound-state calculations
remain useful as reference problems. They isolate the QCD correlation
scale, define the would-be radial eigenvalues of the confining dynamics,
and provide a controlled starting point for studying how an unstable
constituent modifies the analytic structure of the spectrum.

In a previous stable-top Salpeter study~\cite{Zhang:2026top}, the first
four $S$-wave eigenvalues of $t\bar b$, $t\bar c$, and $t\bar q$
configurations were obtained. The radial splittings were of order
$0.2-0.3$~GeV, substantially smaller than the physical top width.
The present work determines the finite-width continuation of that
discrete reference spectrum.

The more general question motivating this work is: for a heavy-light
system containing an unstable heavy quark $Q$ with complex pole mass
$\mu_Q=m_Q-i\Gamma_Q/2$, how do the pole widths and radial splittings
behave in the heavy-quark expansion?  The answer, derived in
Sec.~\ref{sec:general-hq} below, is parametric: the pole widths inherit
$\Gamma_Q$ up to $\mathcal O(\Lambda_{\rm kin}^2/m_Q^2)$ corrections
(here and throughout, $\mathcal O$ denotes the standard asymptotic
notation), while radial
splittings remain $\mathcal O(\Lambda_{\rm rad})$. When $\Gamma_Q\gg\Lambda_{\rm
  rad}$, complex poles can survive but a resolvable radial spectrum
cannot. The top quark, with $\Gamma_t\simeq1.42$~GeV and
$\Lambda_{\rm rad}\sim0.2$--$0.3$~GeV, provides an extreme realization
of this hierarchy.

To test this general result quantitatively, we use the instantaneous
Salpeter framework, which permits a consistent complex-mass insertion
into all mass-dependent operator coefficients and builds directly on the
stable-top reference spectrum of~\cite{Zhang:2026top}.  The complex top
pole mass is inserted consistently into the
constituent energy, projectors, and all mass-dependent coefficients
(constant-width complex-mass prescription for the reference Salpeter
operator), which is distinct from a full gauge-theory complex-mass scheme
or a gauge-complete electroweak calculation. The analysis combines four
ingredients that go beyond a simple comparison of $\Gamma_t$ with the level
spacing $\Delta M$: (i) a general heavy-quark-expansion width-inheritance
relation, realized at the operator level by a biorthogonal
Hellmann--Feynman identity on the non-Hermitian Salpeter eigenproblem;
(ii) a consistent complex-pole mass continuation of the Salpeter operator
$H(m_t-i\Gamma_t/2)$, in which the complex top mass enters $\omega_t$,
the projectors $\Lambda_t^\pm$, and every mass-dependent kernel
coefficient; (iii) a low-pole source-projected spectral response
constructed from the raw biorthogonal residues
$Z_n^{\rm raw}=(s^\dagger R_n)(L_n^\dagger s)/(L_n^\dagger R_n)$, without
positive-normalizing or taking real parts; and (iv) a full
$\Gamma_t$-dependent non-Hermitian re-diagonalization that tracks the
left/right eigenvectors, complex poles, and raw residues across
$\Gamma_t$, instead of freezing the residues at their physical-width
values. In addition, an artificial heavy-mass scaling test confirms the
predicted $m_Q^{-2}$ suppression of the width-inheritance correction.
The central distinction is between a \emph{discrete reference
spectrum} and a \emph{spectrally resolvable response}. The overlap
measure $\mathcal R$ is not treated as a universal experimental boundary.
The source-projected resolvent remains short of a collider prediction
because a definite production operator, nonresonant amplitudes,
hadronization, and detector response are not included. It nevertheless
provides a controlled diagnostic connecting the complex Salpeter poles to
the pole sector of a spectral response. The source-projected analysis
uses the complex poles and their Gaussian-source biorthogonal
residues to examine the resolvability of the low-lying radial
pattern. A normalized positive response is retained only as an
auxiliary tool for testing the influence of different pole weights.
A full width-dependent non-Hermitian
evolution is then used to isolate how the inherited pole width affects
the resolvability of the low-lying radial pattern; the heavy-top-limit
frozen-residue scan is retained only as a quantitative cross-check.

The finite-width dissolution found here has a direct implication for
LHC-oriented searches. After the weak decay of the top constituent,
a putative $t\bar q$ correlation feeds a $Wb\bar q$ final-state
topology, with $q=b,c,u$ denoting the spectator flavor. The subsequent
hadronization may produce open heavy-flavor hadrons; in the $t\bar b$
case, a bottomonium component may also arise through color
rearrangement. The present result provides a benchmark: the stable-top
radial reference poles should not be mapped onto multiple narrow
structures in the reconstructed $Wb\bar q$ spectrum, and any
process-level interpretation of such structures must invoke dynamics
beyond the stable-top radial pole organization.

The distinction from toponium is dynamical: in a $t\bar t$ system,
the Coulombic scale $m_t\alpha_s^2$ can compete with the top width,
whereas in a single-top system with a fixed spectator mass, the radial
scale remains at the sub-GeV level. Recent complex-energy $T$-matrix
and threshold Green-function studies for
toponium~\cite{Tang:2026zhq,Nason:2026oka},
potential-model~\cite{Wang:2024hzd,Fu:2025yft},
QCD sum-rule~\cite{Zhang:2025fdp}, and phenomenological
analyses~\cite{Aguilar-Saavedra:2024mnm,Llanes-Estrada:2024phk,Nason:2025hix,Shao:2025dzw}
address the heavy-heavy case; the present work addresses the
heavy-light case.

The remainder of this paper is organized as follows. Section~\ref{sec:framework}
introduces the finite-width Salpeter framework and the general
unstable-heavy-light heavy-quark expansion. Section~\ref{sec:poles}
presents the complex poles and radial overlap. Section~\ref{sec:response}
presents the source-projected pole response and residue analysis.
Section~\ref{sec:contrast} discusses the spin channel and contrast with
toponium. Section~\ref{sec:implications} summarizes the implications for
LHC-oriented modelling. Section~\ref{sec:conclusion} concludes. Numerical
robustness and non-normal eigenvalue diagnostics are collected in the
appendices.

\section{Finite-width Salpeter framework}\label{sec:framework}

The instantaneous Salpeter framework is used here as a controlled
continuation of the stable-top reference spectrum. The central conclusion
depends on the hierarchy between the inherited top width and the radial
QCD scale, while the detailed potential model determines the reference
pole positions and splittings.

\subsection{Bethe--Salpeter equation and instantaneous reduction}

The four-dimensional Bethe--Salpeter equation for a quark--antiquark
bound state is~\cite{Salpeter:1951sz,Zhang:2026top,Qiao:1996re,Wang:2005qx,Cvetic:2004qg}
\begin{equation}
\chi(P,q)=S_1(p_1)\int\frac{d^4k}{(2\pi)^4}
K(P,q,k)\,\chi(P,k)\,S_2(-p_2),
\label{eq:fulle}
\end{equation}
where $\chi(P,q)$ is the BS wave function, $S_{1,2}$ are the fermion
propagators, and $K$ is the interaction kernel. Introducing
$p_1=\alpha_1P+q$ and $p_2=\alpha_2P-q$ with
$\alpha_i=m_i/(m_1+m_2)$, and adopting the instantaneous approximation
$K(P,q,k)\simeq K(q_\perp,k_\perp)$ in the center-of-mass frame, the
equation reduces to the Salpeter form
\begin{align}
(M-\omega_1-\omega_2)\varphi^{++}
 &=\Lambda_1^+\eta\Lambda_2^+,\nonumber\\
(M+\omega_1+\omega_2)\varphi^{--}
 &=-\Lambda_1^-\eta\Lambda_2^-,\nonumber\\
\varphi^{+-}&=\varphi^{-+}=0,
\label{eq:salpeter}
\end{align}
where $\omega_i=\sqrt{m_i^2+q_\perp^2}$, the three-dimensional wave
function is $\varphi(q_\perp)=i\int dq_P/(2\pi)\,\chi(P,q)$, and the
interaction convolution is
\begin{equation}
\eta(q_\perp)=\int\frac{d^3k_\perp}{(2\pi)^3}
V(q_\perp,k_\perp)\,\varphi(k_\perp).
\label{eq:etaconv}
\end{equation}

\subsection{Energy projections, channel wave functions, and constraints}

The projected Salpeter wave functions are
\begin{equation}
\varphi^{\pm\pm}(q_\perp)
=\Lambda_1^\pm(q_\perp)\frac{\slashed P}{M}
\varphi(q_\perp)\frac{\slashed P}{M}\Lambda_2^\pm(q_\perp),
\end{equation}
with the energy projection operators
\begin{equation}
\Lambda_i^\pm(q_\perp)=\frac{1}{2\omega_i}
\left[\frac{\slashed P}{M}\omega_i
\pm(\slashed p_{i\perp}+\eta_i m_i)\right],
\end{equation}
where $\eta_i=+1$ for the quark and $\eta_i=-1$ for the antiquark.

For the $0^-$ ($^1S_0$) pseudoscalar channel, the Salpeter wave function
decomposes as
\begin{equation}
\varphi_{0^-}(q_\perp)=\bigl[Mf_1(q)+\slashed Pf_2(q)
+\slashed q_\perp f_3(q)+\frac{\slashed P\slashed q_\perp}{M}f_4(q)
\bigr]\gamma_5,
\end{equation}
with four scalar functions $f_1,\dots,f_4$. The Salpeter constraints
$\varphi^{+-}=\varphi^{-+}=0$ give
\begin{align}
f_3(q)&=\frac{M(\omega_2-\omega_1)}{m_1\omega_2+m_2\omega_1}f_2(q),\nonumber\\
f_4(q)&=-\frac{M(\omega_1+\omega_2)}{m_1\omega_2+m_2\omega_1}f_1(q),
\label{eq:constraints}
\end{align}
leaving $f_1(q)$ and $f_2(q)$ as the two independent radial amplitudes.
For the $1^-$ ($^3S_1$) vector channel, the wave function is expanded as
\begin{align}
\varphi_{1^-}(q_\perp)&=(q_\perp\!\cdot\!\varepsilon)
\Bigl[f_1(q)+\frac{\slashed P}{M}f_2(q)
+\frac{\slashed q_\perp}{M}f_3(q)
+\frac{\slashed P\slashed q_\perp}{M^2}f_4(q)\Bigr]\nonumber\\
&\quad+M\slashed\varepsilon f_5(q)
+\slashed\varepsilon\slashed Pf_6(q)
+\frac{\slashed P\slashed\varepsilon\slashed q_\perp}{M}f_7(q)
+\frac{(\slashed P\slashed\varepsilon-\slashed\varepsilon\slashed P)}{2M}f_8(q).
\end{align}
The Salpeter constraints reduce the number of independent radial
functions analogously. The Salpeter normalization condition is
\begin{align}
\int\frac{d^3q_\perp}{(2\pi)^3}
{\rm Tr}\Bigl[
&\bar\varphi^{++}\frac{\slashed P}{M}\varphi^{++}
\frac{\slashed P}{M}
-\bar\varphi^{--}\frac{\slashed P}{M}\varphi^{--}
\frac{\slashed P}{M}\Bigr]=2M.
\label{eq:norm}
\end{align}

\subsection{Interaction kernel and numerical realization}

The interaction kernel is the color-screened Cornell potential
\cite{Eichten:1975bh,Stack:1983cw,Otto:1984zza,Barkai:1984ca,Huntley:1986de,Michael:1985rh,
Qiao:1996re,Laermann:1986pu}, with scalar confinement and vector
one-gluon-exchange contributions
\begin{equation}
V=V_S+\gamma^0\otimes\gamma^0 V_V,
\end{equation}
\begin{align}
V_S(r)&=\lambda r\frac{1-e^{-\alpha r}}{\alpha r},\qquad
V_V(r)=-\frac{4}{3}\frac{\alpha_s}{r}e^{-\alpha r}.
\end{align}
We use $m_b=4.96$~GeV, $m_c=1.62$~GeV, $m_u=0.305$~GeV,
$m_t=172.7$~GeV, $\alpha_s=0.11$, $\alpha=0.06$~GeV, and
$\lambda=0.18$~GeV$^2$~\cite{Zhang:2026top}. The momentum grid has
$n_1=501$ points, $\Delta q=0.008$~GeV, and $q_{\max}=4.0$~GeV.
Parameter variations (Appendix~\ref{app:numerics}) show that moderate
changes of the screened Cornell interaction shift the radial splittings
but do not bring them close to the physical top width.
The absolute reference eigenvalues are model dependent and are not
interpreted as physical resonance masses. The conclusions below
depend on the radial separations and their finite-width continuation,
not on the absolute threshold placement. A complete treatment would
include the constituent weak decay $t\to Wb$ and the corresponding
$Wb\bar q$ continuum, which can generate additional shifts and widths.
The color-screened potential used here discretizes the continuum
spectrum onto the finite momentum grid, so the ``reference poles''
should be understood as well-defined mathematical poles of the
correlation problem with a complex top mass. The radial splittings
$\Delta M$, however, are generated by the confining dynamics at the
sub-GeV scale and are insensitive to the overall threshold offset; the
resolvability conclusion depends only on the hierarchy
$\Gamma_t\gg\Delta M$, not on the absolute pole positions.

After using the Salpeter constraints to eliminate dependent radial
amplitudes and discretizing the remaining coupled integral equations,
the problem can be written as a generalized matrix eigenvalue equation.
The resulting matrix has the block form
$H=\left(\begin{smallmatrix}0&B\\C&0\end{smallmatrix}\right)$ with
$B\neq C^T$, and it is neither Hermitian nor normal in the standard
Euclidean inner product. The eigenvalue problem is solved by standard
numerical diagonalization; after the complex-mass replacement, the matrix
becomes non-Hermitian and left and right eigenvectors are required.

\subsection{Complex pole mass and domain of the approximation}

Near the pole of the dressed top propagator, the leading constant-width
approximation is the complex-pole mass prescription~\cite{Denner:2006ic}
\begin{equation}
 m_t\longrightarrow \mu_t=m_t-\frac{i}{2}\Gamma_t.
\label{eq:complexmass}
\end{equation}
We insert the same $\mu_t$ in $\omega_t(q)=\sqrt{\mu_t^2+q^2}$, the
projection operators $\Lambda_t^\pm$, and every mass-dependent coefficient
of the discretized Salpeter operator. All mass-dependent coefficients
entering the constraints and radial kernels are analytically continued
through the same replacement. The square-root branch is continued from
$\Gamma_t=0$ with $\operatorname{Re}\omega_t>0$ and
$\operatorname{Im}\omega_t<0$. This constant-width complex-mass
prescription for the reference Salpeter operator retains the pole part of
the unstable propagator but omits the energy dependence of
$\Sigma_t$, explicit $Wb\bar q$ continuum channels, nonresonant
weak-decay amplitudes, and gauge-complete production and decay matrix
elements~\cite{Nason:2026oka}. It is therefore used as a complex-pole
mass continuation of a bound-state reference operator, not as a
stand-alone gauge-invariant prediction for an LHC cross section. The small ratios
\begin{equation}
\frac{\Gamma_t}{m_t}\simeq8.2\times10^{-3},\qquad
\frac{\Lambda_{\rm QCD}}{m_t}\sim2\times10^{-3}
\end{equation}
make the pole-mass expansion controlled even though
$\Delta M/\Gamma_t\simeq0.15$--$0.23$ prevents radial spectroscopy.

The complex eigenvalues are written as
\begin{equation}
\Mcal_n=M_n-\frac{i}{2}\Gamma_n.
\label{eq:poles}
\end{equation}

\subsection{Biorthogonal mass derivative and width inheritance}

The discretized Salpeter operator is non-normal~\cite{Trefethen:2005}, so
its right and left eigenvectors are defined by
\begin{equation}
H(\mu_t)R_n=\Mcal_n R_n,\qquad
L_n^\dagger H(\mu_t)=\Mcal_n L_n^\dagger,
\qquad L_n^\dagger R_n=1.
\label{eq:biorthogonal}
\end{equation}
Differentiation with respect to the complex top mass gives the exact
biorthogonal Hellmann--Feynman identity~\cite{Hajong:2024owo}
\begin{equation}
\frac{\partial\Mcal_n}{\partial\mu_t}
=L_n^\dagger\frac{\partial H}{\partial\mu_t}R_n.
\label{eq:nhhf}
\end{equation}
Analytic continuation from $\mu_t=m_t-i\Gamma_t/2$ gives
\begin{align}
\Mcal_n(m_t-i\Gamma_t/2)
={}&M_n(m_t)-\frac{i\Gamma_t}{2}
\left.\frac{\partial M_n}{\partial m_t}\right|_{\Gamma_t=0}\nonumber\\
&-\frac{\Gamma_t^2}{8}
\left.\frac{\partial^2 M_n}{\partial m_t^2}\right|_{\Gamma_t=0}
+\mathcal O(\Gamma_t^3/m_t^2).
\label{eq:linearwidth}
\end{align}
The quadratic term is real for the stable operator, so the pole width
obeys
\begin{equation}
\Gamma_n=\Gamma_t
\left.\frac{\partial M_n}{\partial m_t}\right|_{\Gamma_t=0}
+\mathcal O(\Gamma_t^3/m_t^2).
\label{eq:widthinherit}
\end{equation}
For a single-top system with fixed spectator mass, the heavy-top
expansion $M_n=m_t+E_n^{(0)}+\mathcal O(\Lambda_{\rm kin}^2/m_t)$
implies
\begin{equation}
\Gamma_n=\Gamma_t\left[1+
\mathcal O\!\left(\frac{\Lambda_{\rm kin}^2}{m_t^2}\right)
\right].
\label{eq:widthpower}
\end{equation}
The dominant kinetic part of the mass derivative can be estimated from
the normalized stable-limit radial weight as
\begin{equation}
\Gamma_n^{\rm kin}=\Gamma_t
\left\langle\frac{m_t}{\sqrt{m_t^2+q^2}}\right\rangle_n^{\rm st}
=\Gamma_t\left[1-\frac{\langle q^2\rangle_n^{\rm st}}{2m_t^2}
+\mathcal O\!\left(\frac{q^4}{m_t^4}\right)\right].
\label{eq:kinwidth}
\end{equation}
This kinetic expression is an auxiliary estimator; the exact
biorthogonal identity in Eq.~\eqref{eq:nhhf} is the fundamental relation.
The complete complex-matrix calculation below provides the numerical test
of Eqs.~\eqref{eq:widthinherit} and \eqref{eq:widthpower}.

A symmetric finite-difference evaluation of the stable-spectrum mass
derivative $\partial M_n/\partial m_t$ at $\Gamma_t=0$ gives values
within a few $10^{-5}$ of unity for the four lowest $t\bar b$ radial states.
Multiplying these derivatives by the physical top width reproduces the
full complex-eigenvalue widths $\Gamma_n^{\rm full}$ at the same level,
with relative differences below $10^{-5}$ in units of $\Gamma_t$,
providing a direct numerical verification of the biorthogonal
width-inheritance relation in Eq.~\eqref{eq:widthinherit}. The state-by-state
comparison is summarized in Table~\ref{tab:hf-verify}; detailed
$\delta m$ stability tests are collected in Appendix~\ref{app:numerics}.

\begin{table}[htbp]
\centering
\caption{Numerical verification of the biorthogonal Hellmann--Feynman
width-inheritance relation for the four lowest $t\bar b$ $0^-$ radial
states. The full stable-eigenvalue mass derivative is evaluated at the
stable point $\Gamma_t=0$ by symmetric finite difference with step
$\delta m=10^{-4}m_t$ and multiplied by the physical top width
$\Gamma_t=1.42$~GeV; the result is compared with the full complex-eigenvalue
width $\Gamma_n^{\rm full}$. The derivatives are evaluated with the same
Hamiltonian and finite-difference pipeline as the heavy-mass scaling
test (Table~\ref{tab:scaling}).}
\label{tab:hf-verify}
\begin{tabular}{|c|c|c|c|c|}
\hline
State & $\partial M_n/\partial m_t$ & $\Gamma_t(\partial M_n/\partial m_t)$
& $\Gamma_n^{\rm full}$ (GeV) & $|\Delta|/\Gamma_t$\\
\hline
$1S$ & $0.99998$ & $1.41997$ & $1.41997$ & $<10^{-5}$\\
$2S$ & $0.99997$ & $1.41996$ & $1.41995$ & $<10^{-5}$\\
$3S$ & $0.99996$ & $1.41994$ & $1.41994$ & $<10^{-5}$\\
$4S$ & $0.99995$ & $1.41993$ & $1.41993$ & $<10^{-5}$\\
\hline
\end{tabular}
\end{table}
The agreement demonstrates that the near-universal pole width is
generated by the mass sensitivity of the complete Salpeter eigenvalue
rather than imposed as an external broadening prescription.

\subsection{Overlap measure and an analytic double-peak test}

For adjacent poles we define
\begin{equation}
\mathcal R_n=\frac{M_{n+1}-M_n}{(\Gamma_n+\Gamma_{n+1})/2}.
\label{eq:R}
\end{equation}
Values $\mathcal R\gg1$ indicate parametrically separated poles, while
$\mathcal R\ll1$ indicates strong overlap. A useful analytic benchmark
is provided by two positive Lorentzians of equal width $\Gamma$, equal
strength, and separation $\Delta M$. The midpoint changes from a maximum
to a minimum when
\begin{equation}
\frac{\Delta M}{\Gamma}>\frac{1}{\sqrt3}.
\label{eq:twopeak}
\end{equation}
Thus even the most symmetric two-pole problem does not use $\mathcal R=1$
as an exact crossover. In the present systems, the computed $\mathcal R$
values are so far below $1/\sqrt3$ that the conclusion is not sensitive
to this ambiguity.

\subsection{General unstable-heavy-light limit}
\label{sec:general-hq}

Before specializing to the top quark, it is useful to state the general
heavy-mass-expansion result for a system containing one unstable heavy
constituent with complex pole mass $\mu_Q=m_Q-i\Gamma_Q/2$ and a fixed
spectator of mass $m_s$.  The stable ($\Gamma_Q=0$) poles admit the
expansion
\begin{equation}
\Mcal_n(\mu_Q;m_s)
=
\mu_Q
+
E_n^{(0)}(m_s)
+
\frac{C_n(m_s)}{\mu_Q}
+
\mathcal O\!\left(\frac{\Lambda_{\rm kin}^3}{m_Q^2}\right),
\label{eq:general-hq-pole}
\end{equation}
where $E_n^{(0)}(m_s)$ includes the spectator rest mass and static
correlation energy (not necessarily $\mathcal O(\Lambda_{\rm QCD})$),
$C_n(m_s)=\mathcal O(\Lambda_{\rm kin}^2)$ with $\Lambda_{\rm kin}^2\sim\langle
q^2\rangle$ the kinetic scale, and $\Lambda_{\rm rad}\sim\Delta M_n$ is
the radial excitation scale.  Expanding $1/\mu_Q$ in powers of
$\Gamma_Q/m_Q$ and collecting the imaginary part gives
\begin{equation}
\Gamma_n
=
\Gamma_Q
\left[
1
+
\mathcal O\!\left(\frac{\Lambda_{\rm kin}^2}{m_Q^2}\right)
\right].
\label{eq:general-gamma}
\end{equation}
The radial splittings, by contrast, satisfy
\begin{equation}
\Delta M_n
=
E_{n+1}^{(0)}-E_n^{(0)}
+
\mathcal O\!\left(\frac{\Lambda_{\rm kin}^2}{m_Q}\right)
=
\mathcal O(\Lambda_{\rm rad}),
\label{eq:general-deltam}
\end{equation}
which do not grow linearly with $m_Q$.  Since $\Delta E_n^{(0)}=\mathcal O(\Lambda_{\rm rad})$, the relative correction to the
splitting is generically $\mathcal O(\Lambda_{\rm kin}^2/(m_Q\,\Lambda_{\rm rad}))$. Consequently,
\begin{equation}
\frac{\Delta M_n}{\Gamma_n}
=
\frac{\Delta E_n^{(0)}}{\Gamma_Q}
\left[
1+\mathcal O\!\left(\frac{\Lambda_{\rm kin}^2}{m_Q\,\Lambda_{\rm rad}}\right)
\right].
\label{eq:general-ratio}
\end{equation}
Within the fixed-spectator heavy-mass expansion and the constant
complex-pole-mass continuation, an unstable system with one heavy
constituent has a parametrically simple structure: the pole width is
inherited from the unstable heavy constituent up to
$\mathcal O(\Lambda_{\rm kin}^2/m_Q^2)$ corrections, whereas radial splittings
remain $\mathcal O(\Lambda_{\rm rad})$.  Therefore, if
$\Gamma_Q\gg\Lambda_{\rm rad}$, the existence of complex poles does not
imply a resolvable radial spectrum.  This leading-power result is
independent of the detailed Salpeter kernel within the fixed-spectator
heavy-mass expansion, conditional on the constant complex-pole-mass
prescription.

The biorthogonal Hellmann--Feynman identity of
Eq.~\eqref{eq:nhhf} provides the operator-level realization of this
general result: $\partial\Mcal_n/\partial\mu_Q$ is computed directly from
the left and right eigenvectors, and the numerical result
$\Gamma_n/\Gamma_t=0.99995$--$1.00000$ serves as a quantitative test of
Eq.~\eqref{eq:general-gamma} for $m_Q=m_t$, not as the source of the
relation. An artificial heavy-mass scaling test
(Sec.~\ref{sec:scaling}) verifies the predicted $m_Q^{-2}$
suppression of $|1-\partial M_n/\partial m_Q|$.

For the top quark, $\Gamma_t\simeq1.42$~GeV and
$\Lambda_{\rm rad}\sim0.2$--$0.3$~GeV, so $\Gamma_t\gg\Lambda_{\rm rad}$
and the top is the extreme realization of this hierarchy.

\section{Complex poles and radial overlap}\label{sec:poles}

\begin{table}[htbp]
\centering
\caption{Complex $0^-$ $t\bar b$ poles at $\Gamma_t=1.42$~GeV. Stable
masses $M_n^{(0)}$ are at $\Gamma_t=0$; $\mathcal R_n$ refers to the
overlap measure defined in Eq.~\eqref{eq:R}.}
\label{tab:tbpoles}
\begin{tabular}{|c|c|c|c|c|c|}
\hline
State & $M_n^{(0)}$ (GeV) & $M_n$ (GeV) & $\Gamma_n$ (GeV) & $\Delta M_n$ (GeV) & $\mathcal R_n$\\
\hline
$1S$ & 177.837 & 177.837 & 1.4200 & 0.293 & 0.207\\
$2S$ & 178.130 & 178.130 & 1.4200 & 0.202 & 0.142\\
$3S$ & 178.332 & 178.332 & 1.4199 & 0.163 & 0.115\\
$4S$ & 178.495 & 178.495 & 1.4199 & -- & --\\
\hline
\end{tabular}
\end{table}

Table~\ref{tab:tbpoles} shows that the physical width is inherited by
all four radial poles, while the real parts are essentially unchanged
from the stable limit. The full widths and the leading kinetic estimate
of Eq.~\eqref{eq:kinwidth} agree within several $10^{-5}$ in units of
$\Gamma_t$ (Appendix~\ref{app:numerics}).

\begin{table}[htbp]
\centering
\caption{First radial spacing, overlap measure, and survival estimate
for the three single-top channels.}
\label{tab:channels}
\begin{tabular}{|c|c|c|c|c|}
\hline
Channel & $M_{1S}$ (GeV) & $\Delta M_{2S-1S}$ (GeV) & $\mathcal R_1$ & $e^{-1/\mathcal R_1}$\\
\hline
$t\bar b$ & 177.837 & 0.293 & 0.207 & $\sim10^{-2}$\\
$t\bar c$ & 174.662 & 0.327 & 0.230 & $\sim10^{-2}$\\
$t\bar u$ & 173.437 & 0.314 & 0.221 & $\sim10^{-2}$\\
\hline
\end{tabular}
\end{table}

All first radial splittings satisfy $\mathcal R_1<0.24$, and the higher
$t\bar b$ splittings are smaller still. They are well below the
equal-Lorentzian double-maximum threshold in Eq.~\eqref{eq:twopeak}.
Potential variations, the experimental top width range, and grid tests
leave this hierarchy unchanged (Appendix~\ref{app:numerics}).

\subsection{Heavy-mass scaling test}
\label{sec:scaling}

\begin{figure}[H]
\centering
\includegraphics[width=0.82\textwidth]{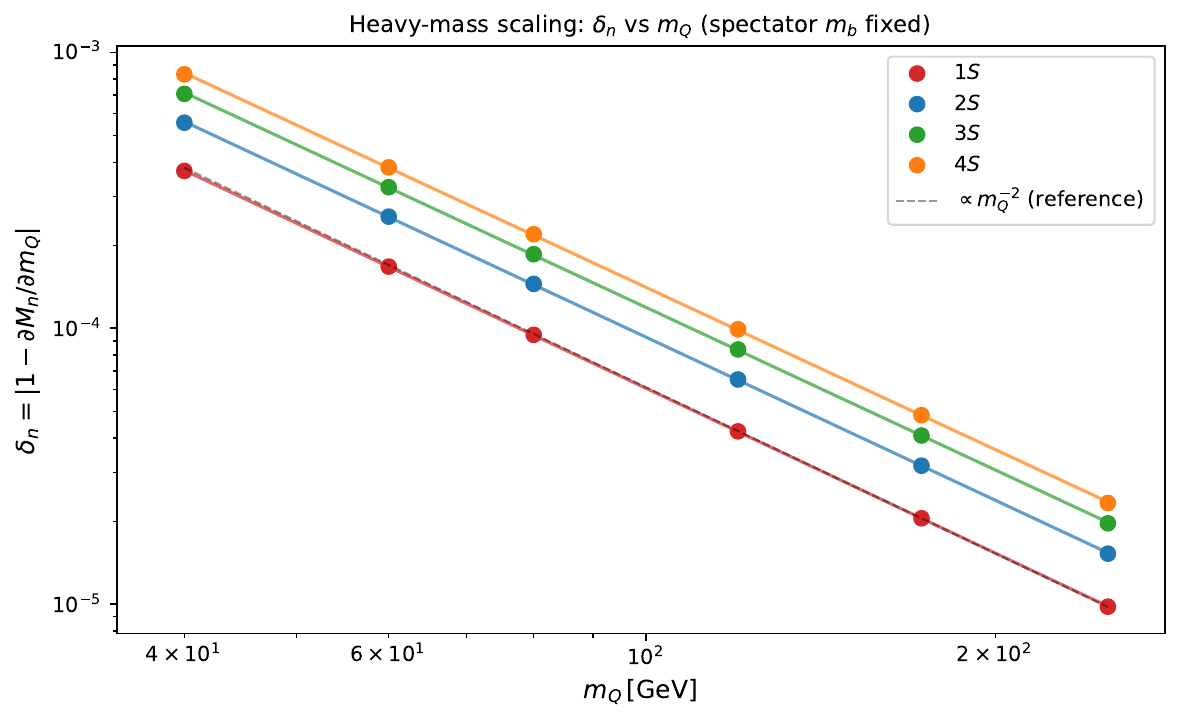}
\caption{Heavy-mass scaling test: $\delta_n=|1-\partial M_n/\partial m_Q|$
vs.\ $m_Q$ for the four lowest $t\bar b$-type radial states (spectator
$m_b=4.96$~GeV fixed, kernel parameters fixed). Dashed line: reference
$\propto m_Q^{-2}$. The fitted exponents are $p_n\simeq1.95$--$1.99$
with $R^2>0.999$ for all states. This test is designed to validate the
heavy-mass power counting rather than to represent additional physical
quark species.}
\label{fig:heavy-scaling}
\end{figure}

To verify the $\mathcal O(\Lambda_{\rm kin}^2/m_Q^2)$ suppression predicted by
Eq.~\eqref{eq:general-gamma}, we perform an artificial heavy-mass scan:
the heavy-quark mass is varied as $m_Q=40$, $60$, $80$, $120$, $172.7$,
$250$~GeV while the spectator mass ($m_b=4.96$~GeV), kernel parameters,
and source definition are held fixed. For each $m_Q$ the stable
($\Gamma_Q=0$) Salpeter eigenvalues and their symmetric finite-difference
mass derivatives $D_n(m_Q)\equiv\partial M_n/\partial m_Q$ are computed.
The deviation
\begin{equation}
\delta_n(m_Q)=\left|1-D_n(m_Q)\right|
\label{eq:delta-def}
\end{equation}
measures the departure of the width-inheritance relation from exactness.
Figure~\ref{fig:heavy-scaling} shows a log-log plot of $\delta_n$ versus
$m_Q$ for the four lowest radial states, together with a reference
$\propto m_Q^{-2}$ slope.

\begin{table}[htbp]
\centering
\caption{Fitted power-law exponents $p_n$ for
$\delta_n=A_n\,m_Q^{-p_n}$ and $\delta_n$ at $m_Q=172.7$~GeV.
Expectation: $p_n\simeq2$ from the heavy-quark expansion.}
\label{tab:scaling}
\begin{tabular}{|c|c|c|c|c|}
\hline
State & fitted $p_n$ & $\delta_n(172.7~\mathrm{GeV})$ & $R^2$ & expectation\\
\hline
$1S$ & $1.99$ & $2.05\times10^{-5}$ & $1.0000$ & $m_Q^{-2}$\\
$2S$ & $1.96$ & $3.18\times10^{-5}$ & $1.0000$ & $m_Q^{-2}$\\
$3S$ & $1.96$ & $4.09\times10^{-5}$ & $1.0000$ & $m_Q^{-2}$\\
$4S$ & $1.95$ & $4.84\times10^{-5}$ & $1.0000$ & $m_Q^{-2}$\\
\hline
\end{tabular}
\end{table}

The artificial heavy-mass scan provides a direct numerical validation
of the $m_Q^{-2}$ suppression predicted by the fixed-spectator
heavy-mass expansion. The fitted exponents $p_n\simeq1.95$--$1.99$
are consistent with the expected scaling within the fit uncertainty,
confirming that the width-inheritance relation $\Gamma_n\simeq\Gamma_Q$
is controlled by the heavy-quark expansion and its correction is
parametrically suppressed as $\Lambda_{\rm kin}^2/m_Q^2$. At
$m_Q=172.7$~GeV the derivatives used in the scaling analysis reproduce
those in Table~\ref{tab:hf-verify}, since both are evaluated with the
same Hamiltonian and finite-difference pipeline.

As a complementary check, the fixed-spectator residual energies
$E_n^{\rm res}(m_Q)=M_n(m_Q)-m_Q$ are nearly $m_Q$-independent across
the scan:
\begin{center}
\begin{tabular}{|c|c|c|c|c|}
\hline
$m_Q$ (GeV) & $E_{1S}^{\rm res}$ & $E_{2S}^{\rm res}$ &
$E_{3S}^{\rm res}$ & $E_{4S}^{\rm res}$\\
\hline
$40$   & $5.148$ & $5.448$ & $5.654$ & $5.821$\\
$80$   & $5.141$ & $5.436$ & $5.640$ & $5.804$\\
$172.7$ & $5.137$ & $5.430$ & $5.632$ & $5.795$\\
$250$  & $5.136$ & $5.428$ & $5.630$ & $5.792$\\
\hline
\end{tabular}
\end{center}
The variation is below $0.3\%$ across $m_Q=40$--$250$~GeV, showing
that the fixed-spectator residual energies remain nearly independent
of $m_Q$, as expected in the heavy-mass limit. Their radial
differences, $E_{n+1}^{\rm res}-E_n^{\rm res}$, remain of
$\mathcal O(\Lambda_{\rm rad})$, confirming that the radial excitation scale
does not grow with the heavy-quark mass, as predicted by
Eq.~\eqref{eq:general-deltam}.

\section{Source-projected pole response}\label{sec:response}

While the complex poles obtained in Sec.~III determine
the positions and widths of the radial modes, they do
not by themselves determine whether multiple maxima
appear in an energy-dependent response. This also
depends on how strongly a given source couples to
each mode. We therefore examine the resolvent
projected onto a smooth momentum-space source,
for which the pole contributions are weighted by
the corresponding biorthogonal residues. The aim
is to determine whether the radial structure remains
resolvable as the top-quark width increases, taking
into account the unequal and source-dependent
strengths of the radial poles. This construction
provides a diagnostic of the pole sector of the
Salpeter problem rather than a prediction for a
specific collider production process.

\subsection{Gaussian source and biorthogonal basis}

The source vector $s$ is a Gaussian momentum-space wave packet with
width $\beta=0.5$~GeV,
\begin{equation}
s_a(q_i)={\cal N}_\beta\,c_a\,
\exp\!\left(-\frac{q_i^2}{2\beta^2}\right),
\label{eq:source}
\end{equation}
where $q_i$ is the $i$-th discrete momentum grid point, $a$ indexes the
independent Salpeter radial amplitudes, $c_a$ specifies the component
weights, and ${\cal N}_\beta$ is a normalization constant such that
$s^\dagger s=1$. The source is constructed with the $0^-$ Salpeter
projection structure, coupling to both independent radial components
$f_1(q)$ and $f_2(q)$ with the same Gaussian envelope; we choose
$c_1=c_2=1$ before overall normalization. The Gaussian source is
introduced as a smooth diagnostic probe of the low-momentum Salpeter
subspace and is not identified with a unique collider production current
or with a physical electroweak current. Smaller
$\beta$ emphasizes the very-low-momentum region, whereas larger $\beta$
samples a broader momentum range and corresponds to a more localized
coordinate-space probe. The left and right source vectors are taken to
have the same component profile in the discretized basis.

The quadrature factors from the momentum-space integration are absorbed
into the discretized basis vectors, so the biorthogonal products in the
following equations are ordinary Euclidean products in the weighted
basis. The numerical implementation maintains
$L_m^\dagger R_n=\delta_{mn}$ to better than $10^{-13}$ for the
low-lying eigenpairs.

\subsection{Resolvent and pole residues}\label{sec:resolvent-recon}

For the source vector $s$ in the discretized Salpeter space, the
projected resolvent is
\begin{equation}
G(E)=s^\dagger\frac{1}{E-H}s
=\sum_{n\in{\cal S}}\frac{Z_n^{\rm raw}}{E-\Mcal_n}
+G_{\rm rem}^{({\cal S})}(E),
\label{eq:green}
\end{equation}
with the exact biorthogonal pole residues
\begin{equation}
Z_n^{\rm raw}
=\frac{(s^\dagger R_n)(L_n^\dagger s)}{L_n^\dagger R_n}.
\label{eq:source-residue}
\end{equation}
Here ${\cal S}$ denotes a selected set of low-lying poles.
For the four-pole analysis below,
${\cal S}=\{1S,2S,3S,4S\}$.
Within the finite Salpeter basis,
$G_{\rm rem}^{({\cal S})}$ denotes the contribution of
eigenmodes omitted from this low-pole set. Physical continuum
contributions are not contained in the present
finite-dimensional bound-state resolvent and would have to be
incorporated separately in a process-level correlation
function. The Gaussian source defines a controlled resolution
probe of the bound-state subspace rather than a collider production
operator, and the residues $Z_n^{\rm raw}$ are therefore
source-dependent Salpeter residues rather than universal
physical couplings.

The \emph{low-pole source-projected spectral response} is defined as
\begin{equation}
\rho_G^{(N)}(E)=-\frac{1}{\pi}\operatorname{Im}G_N(E)
=-\frac{1}{\pi}\operatorname{Im}
\sum_{n=1}^{N}\frac{Z_n^{\rm raw}}{E-\Mcal_n},
\label{eq:rhoG}
\end{equation}
where $N=|{\cal S}|$ and $Z_n^{\rm raw}$ is the complex biorthogonal
residue of Eq.~\eqref{eq:source-residue}. The sum is truncated at
$N$ physical poles; higher modes enter through
$G_{\rm rem}^{({\cal S})}$. This quantity is the primary spectral
diagnostic of the present work: it uses the raw residues without
taking real parts, positive-normalizing, or sum-normalizing. A
truncated non-Hermitian pole-sector response is not guaranteed to be
globally positive, and the distinction between the raw low-pole response and
the normalized positive response discussed below is exactly the
distinction between a pole-sector spectral diagnostic and a positive
visualization aid. We therefore use the response constructed from the
full complex biorthogonal residues as the primary spectral diagnostic.
The normalized positive response is retained as an auxiliary quantity
for examining the influence of different relative pole weights.

For the relative comparison of radial weights across channels and
across source parameters, we also use the normalized residues
\begin{equation}
\widehat Z_n=\frac{\operatorname{Re}Z_n^{\rm raw}}
{\sum_{k=1}^{4}\operatorname{Re}Z_k^{\rm raw}},
\label{eq:Znorm}
\end{equation}
which are quoted in Table~\ref{tab:residues-raw} and used for the
auxiliary positive-response analysis, including the residue-weight
scan in Appendix~\ref{app:residue-scan}. The
normalization in Eq.~\eqref{eq:Znorm} is convenient for displaying
the relative structure of the response but does not reproduce the
full-resolvent normalization; the two quantities must not be
interchanged. Because the imaginary parts of the raw residues are
below $2.5\times10^{-5}$ in absolute value (Table~\ref{tab:residues-raw}),
we use their real parts to define the positive pole response, but
the raw complex residues enter $\rho_G^{(N)}(E)$ without any such
projection.

\begin{table}[htbp]
\centering
\caption{Raw biorthogonal Gaussian-source Salpeter residues
$Z_n^{\rm raw}=\operatorname{Re}Z_n^{\rm raw}+i\,\operatorname{Im}Z_n^{\rm raw}$
[Eq.~\eqref{eq:source-residue}] for the three single-top channels at
$\beta=0.5$~GeV, $c_2/c_1=1$, and $\Gamma_t=1.42$~GeV. The imaginary parts
are below $2.5\times10^{-5}$ in absolute value for all channels; the
real parts dominate and are used to form the normalized residues
$\widehat Z_n=\operatorname{Re}Z_n^{\rm raw}/\sum_k\operatorname{Re}Z_k^{\rm raw}$
quoted in the last four columns.}
\label{tab:residues-raw}
\begin{tabular}{|c|c|c|c|c|c|c|c|c|}
\hline
Channel & $\operatorname{Re}Z_{1S}^{\rm raw}$ & $\operatorname{Re}Z_{2S}^{\rm raw}$
& $\operatorname{Re}Z_{3S}^{\rm raw}$ & $\operatorname{Re}Z_{4S}^{\rm raw}$
& $\widehat Z_{1S}$ & $\widehat Z_{2S}$ & $\widehat Z_{3S}$ & $\widehat Z_{4S}$\\
\hline
$t\bar b$ & 0.139030 & 0.324620 & 0.253004 & 0.159529
& 0.159 & 0.370 & 0.289 & 0.182\\
$t\bar c$ & 0.358122 & 0.421413 & 0.153491 & 0.053467
& 0.363 & 0.427 & 0.156 & 0.054\\
$t\bar u$ & 0.737144 & 0.169078 & 0.004453 & 0.037061
& 0.778 & 0.178 & 0.005 & 0.039\\
\hline
\end{tabular}
\end{table}

Table~\ref{tab:residues-raw} lists the raw biorthogonal residues
$Z_n^{\rm raw}$ for the three channels. The real parts dominate the
imaginary parts by more than four orders of magnitude, justifying the
use of $\operatorname{Re}Z_n^{\rm raw}$ in forming the positive pole
response of Sec.~\ref{sec:pos-response}. The residue patterns differ
significantly across the three channels. The radial Salpeter wave
functions of the corresponding stable-top systems were presented
in Ref.~\cite{Zhang:2026top} (Fig.~1 and Appendix~A, Figs.~A1
and~A4). These wave functions show that the momentum distributions
and radial node positions depend on the spectator-quark mass. For
the Gaussian source defined in Eq.~\eqref{eq:source}, the overlap
factors entering Eq.~\eqref{eq:source-residue} sample both radial
components over a finite momentum range. The sign changes across
radial nodes can produce partial cancellations in these overlaps,
with the extent of cancellation depending on the radial state and
spectator mass. Consequently, the residues need not decrease
monotonically with increasing radial excitation.

For the source parameters used in Table~\ref{tab:residues-raw}, the
$t\bar b$ and $t\bar c$ channels receive substantial contributions
from excited radial states, whereas the $t\bar u$ channel is strongly
dominated by the $1S$ contribution. The previously published wave
functions provide a qualitative understanding of these patterns,
while the residues reported here are calculated directly from the
biorthogonal eigenvectors of the finite-width Salpeter operator. The
biorthogonality relation $L_m^\dagger R_n=\delta_{mn}$ is maintained to
better than $10^{-13}$ and the eigen-residual norms
$\|HR_n-\Mcal_nR_n\|_2/|\Mcal_n|$ and
$\|L_n^\dagger H-\Mcal_n L_n^\dagger\|_2/|\Mcal_n|$ are below
$10^{-13}$ for all physical poles and all channels.

\subsection{Physical-width spectral response}\label{sec:pos-response}

\begin{figure}[H]
\centering
\includegraphics[width=0.82\textwidth]{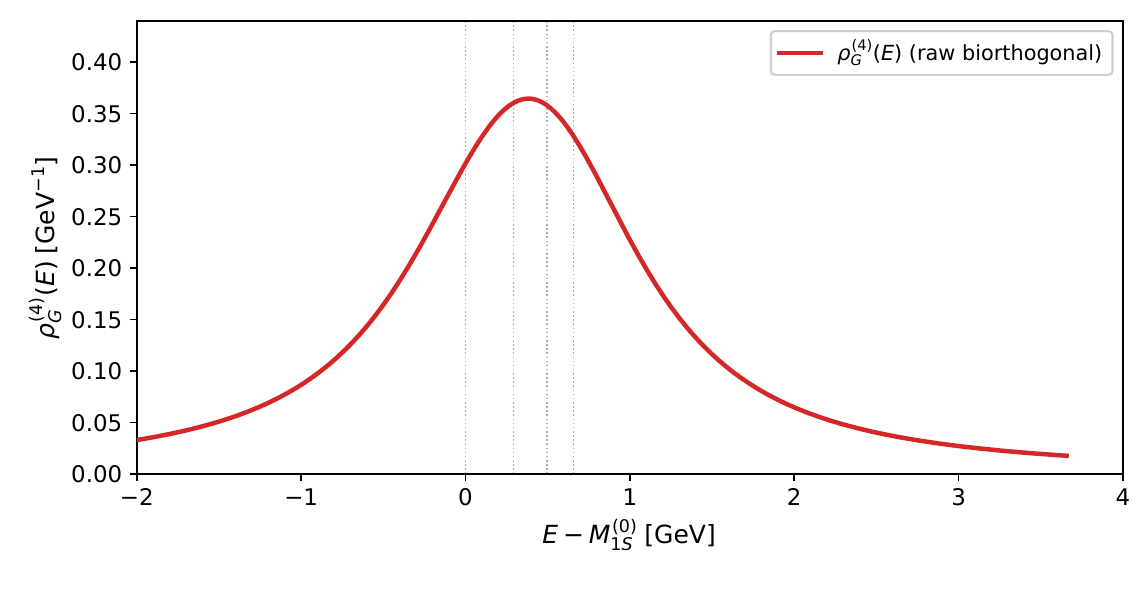}
\caption{Physical-width $t\bar b$ low-pole source-projected
response $\rho_G^{(4)}(E)$ at $\Gamma_t=1.42$~GeV,
$\beta=0.5$~GeV, and $c_2/c_1=1$, calculated using the
full complex biorthogonal residues of Eq.~\eqref{eq:source-residue}.
The vertical lines indicate the stable-top radial
reference positions. Despite the presence of four
complex radial poles, the response exhibits a
single broad maximum.}
\label{fig:pole-response}
\end{figure}

The low-pole source-projected response
$\rho_G^{(4)}(E)=-(1/\pi)\operatorname{Im}\sum_{n=1}^{4}Z_n^{\rm raw}/(E-\Mcal_n)$
of Eq.~\eqref{eq:rhoG} uses the full complex biorthogonal residues defined
in Eq.~\eqref{eq:source-residue}, whose real parts are listed in
Table~\ref{tab:residues-raw}, with no real-part projection or
positive-normalization step. The physical-width response in the $t\bar b$
channel is shown in Fig.~\ref{fig:pole-response}. It has a
single broad maximum at $E\simeq178.22$~GeV, located inside the pole band
$[177.84,178.50]$~GeV and shifted away from any individual pole because
the four contributions overlap strongly. The response is strictly
positive: across the displayed energy window the minimum-to-maximum
ratio is $+4.6\times10^{-2}$ for $t\bar b$, $+4.5\times10^{-2}$ for
$t\bar c$, and $+4.1\times10^{-2}$ for $t\bar u$. No negative excursion
appears over the displayed energy range. The single-maximum structure
is therefore obtained directly from the unnormalized, complex-residue
response, without enforcing positivity through a replacement of the
residues.

For the auxiliary analysis of relative pole weights, we also
define the normalized positive pole response
\begin{equation}
\rho^{(+)}(E)=\sum_{n=1}^{4}\frac{\widehat Z_n}{\pi}
\frac{\Gamma_n/2}{(E-M_n)^2+\Gamma_n^2/4},
\label{eq:rho}
\end{equation}
using the normalized residues $\widehat Z_n$ from
Table~\ref{tab:residues-raw}. The normalized positive response is
related to the raw low-pole response mainly by an overall factor.
Writing
$S_4=\sum_{k=1}^{4}\operatorname{Re}Z_k^{\rm raw}$,
we obtain
\[
\rho_G^{(4)}(E)=S_4\rho^{(+)}(E)
-\frac{1}{\pi}\sum_{n=1}^{4}
\frac{\operatorname{Im}Z_n^{\rm raw}(E-M_n)}
{(E-M_n)^2+\Gamma_n^2/4}.
\]
For the $t\bar b$ channel, Table~\ref{tab:residues-raw} gives
$S_4=0.876183$, so that
$\rho^{(+)}(E)\simeq1.1413\,\rho_G^{(4)}(E)$
when the small imaginary parts of the residues
are neglected. Accordingly, displaying the raw and normalized
positive responses together would not provide an
independent test of radial resolvability.
Figure~\ref{fig:pole-response} therefore displays only the response
constructed from the full complex biorthogonal
residues. The normalized positive response of
Eq.~\eqref{eq:rho} is used solely as an auxiliary tool for
the residue-weight studies in Appendix~\ref{app:residue-scan}.

\begin{figure}[H]
\centering
\includegraphics[width=0.82\textwidth]{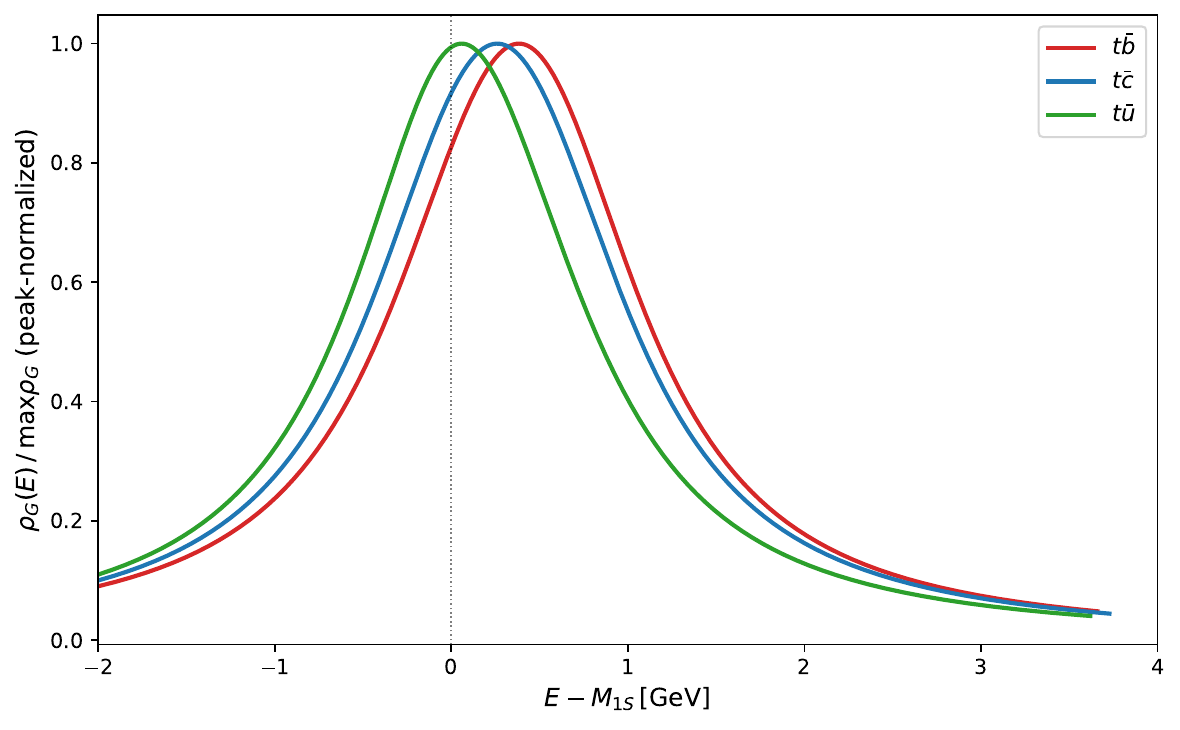}
\caption{Peak-normalized four-pole source-projected responses for the
$t\bar b$, $t\bar c$, and $t\bar u$ channels at $\Gamma_t=1.42$~GeV,
$\beta=0.5$~GeV, and $c_2/c_1=1$. Each curve is divided by its own
maximum $A_\alpha=\max_E\rho_{G,\alpha}^{(4)}(E)$ and plotted against
$E-M_{1S,\alpha}^{(0)}$, where $M_{1S,\alpha}^{(0)}$ is the stable
$1S$ reference mass in that channel. The plotted responses are
dimensionless. All three channels exhibit a single broad maximum,
whose relative position depends on the channel-dependent pole weights
and radial separations.}
\label{fig:three-channel}
\end{figure}

Figure~\ref{fig:three-channel} compares the low-pole source-projected
responses of the three channels at the physical top width, with the
energy measured relative to the stable $1S$ reference mass of each
channel. Although all three responses exhibit a single broad maximum,
the locations of these maxima are different. For $t\bar b$, the
maximum occurs at $E-M_{1S}^{(0)}\simeq0.382$~GeV, as reported in
Table~\ref{tab:peak-count}.

The relative shifts are closely related to the channel-dependent
residue distributions in Table~\ref{tab:residues-raw}. The excited
states ($2S$--$4S$) account for approximately $84\%$ of the
normalized residue weight in the $t\bar b$ channel, giving a
substantial contribution on the higher-energy side of the $1S$ pole.
The corresponding excited-state fraction is about $64\%$ for
$t\bar c$, whereas the $t\bar u$ response is dominated by the $1S$
contribution (approximately $78\%$). Accordingly, the combined
maximum is displaced farthest from the $1S$ reference position for
$t\bar b$ and remains closest to it for $t\bar u$. The precise peak
positions are determined jointly by the residue weights, radial pole
separations, and pole widths.

For each channel $\alpha$, the response plotted in
Fig.~\ref{fig:three-channel} is normalized to its own maximum,
\[
\widetilde{\rho}_{G,\alpha}^{(4)}(E)
=\frac{\rho_{G,\alpha}^{(4)}(E)}
{A_\alpha},\qquad
A_\alpha=\max_E\rho_{G,\alpha}^{(4)}(E).
\]
This peak normalization preserves the peak positions and the number
of maxima but removes the differences in absolute response strength.
It is distinct from the residue normalization introduced in
Eq.~\eqref{eq:Znorm}.

\subsection{Width-driven dissolution}\label{sec:dissolution}

The title of this work emphasizes the loss of radial resolvability as
the inherited top width increases. To establish that this loss is not a
frozen-residue artifact, we perform a \emph{full width-dependent
non-Hermitian re-diagonalization}: for each value of $\Gamma_t$ the
Salpeter operator $H(m_t-i\Gamma_t/2)$ is reconstructed from scratch and
the biorthogonal eigenproblem is solved anew, giving
$\Mcal_n(\Gamma_t)$, $L_n(\Gamma_t)$, $R_n(\Gamma_t)$, and the raw
residues $Z_n^{\rm raw}(\Gamma_t)$. The full-recalculation response is
\begin{equation}
\rho_G(E;\Gamma_t)
=-\frac{1}{\pi}\operatorname{Im}
\sum_{n=1}^{4}
\frac{Z_n^{\rm raw}(\Gamma_t)}
{E-\Mcal_n(\Gamma_t)}.
\label{eq:rhoGgam}
\end{equation}
The state tracking across widths uses the biorthogonal overlap
$\mathcal O_{mn}=|L_m^\dagger(\Gamma_i)R_n(\Gamma_{i+1})|$, with greedy
maximum-overlap assignment; the resulting assignment is the identity
across all transitions, with maximum off-diagonal overlap below
$4\times10^{-5}$ for all three channels.

For direct comparison we also retain the leading heavy-top-limit
\emph{frozen-residue} approximation,
\begin{equation}
\rho_{\rm frozen}(E;\Gamma_t)
=-\frac{1}{\pi}\operatorname{Im}
\sum_{n=1}^{4}
\frac{Z_n^{\rm raw}(\Gamma_t^{\rm phys})}
{E-M_n^{(0)}+i\Gamma_t/2},
\label{eq:rhofrozen}
\end{equation}
which fixes the real pole positions at the stable-limit values
$M_n^{(0)}$, sets $\Gamma_n=\Gamma_t$, and freezes the residues at their
physical-width values. The frozen-residue scan captures the qualitative
dissolution pattern, while the full non-Hermitian evolution modifies
the intermediate-width crossover quantitatively; both prescriptions
agree that the physical-width response has a single broad maximum.

The scan is performed at $\Gamma_t=0$, $0.05$, $0.10$, $0.20$, $0.30$,
$0.50$, $0.70$, $1.00$, and $1.42$~GeV (Table~\ref{tab:peak-count}).
For the heatmap, $\Gamma_t$ is sampled at $N_\Gamma=50$ uniformly
spaced points between $0.05$ and $1.42$~GeV. Only the four lowest
radial poles are retained, so that the figure isolates the
resolvability of the $1S$--$4S$ pattern without incorporating
additional discretized-basis eigenmodes. The $\Gamma_t=0$ limit is
excluded from the heatmap because the Lorentzian peaks approach delta
distributions whose sampled heights become grid dependent. For
visualization, each response is normalized to its own maximum,
\begin{equation}
\widetilde\rho_G(E;\Gamma_t)
=
\frac{
\rho_G(E;\Gamma_t)
}{
\max_E \rho_G(E;\Gamma_t)
}.
\label{eq:rhotilde}
\end{equation}
Each curve is normalized to its own maximum for visualization, so that
the narrow small-width peaks do not compress the physical-width
response. This normalization removes the absolute response strength and
is used only to compare the number and locations of resolvable maxima.

The peak count is determined with \texttt{scipy.signal.find\_peaks},
using a relative prominence threshold of $5\%$ of the curve maximum and a
minimum separation of five grid points. Table~\ref{tab:peak-count}
reports the peak count for the full-recalculation response
$\rho_G(E;\Gamma_t)$ of Eq.~\eqref{eq:rhoGgam}. In the $t\bar b$
channel, four resolved maxima remain visible up to
$\Gamma_t=0.10$~GeV. At $\Gamma_t=0.20$~GeV, the weakest $4S$ feature
falls below the operational prominence threshold, so the four-peak
pattern is no longer fully resolvable. By $\Gamma_t\ge0.30$~GeV, only
one maximum remains. The $t\bar c$ channel shows the same progression
(four peaks at $\Gamma_t\le0.05$~GeV, three at $0.10$~GeV, two at
$0.20$--$0.30$~GeV, one at $\Gamma_t\ge0.50$~GeV). The $t\bar u$ channel
shows only two resolvable peaks at small width because the $1S$
residue is $77.8\%$ of the sum, so the $3S$ and $4S$ features are too
weak to register as separate maxima under the $5\%$ prominence
threshold; the $2S$ feature is clearly visible and merges with $1S$ by
$\Gamma_t=0.20$~GeV. The peak count is operational and should not be
interpreted as a universal experimental boundary. The channel-dependent
peak count at small width illustrates that, at the quantitative level,
the resolvability criterion is source-dependent; however, the
physical-width single-broad-maximum conclusion is robust across all
channels and source variations.

Crucially, the real pole positions move by less than $1.4\times10^{-7}$~GeV
across the entire scan, far smaller than $\Gamma_t$ itself, and the
ratio $\Gamma_n/\Gamma_t$ at the physical width is
$0.99995$--$1.00000$ for all four radial states and all three channels
(Table~\ref{tab:widths-app}). The simple heavy-top-limit width
inheritance $\Gamma_n\simeq\Gamma_t$ is therefore quantitatively
validated by the full non-Hermitian re-diagonalization.

Together with the source-parameter tests in Sec.~\ref{sec:source-dep},
this full evolution illustrates how the hierarchy
$\Gamma_t\gg\Delta M$ removes the resolvability of the four-pole
pattern, and confirms that the conclusion does not rely on a
frozen-residue approximation.

\begin{table}[!ht]
\centering
\caption{Automatic peak count of the full-recalculation response
$\rho_G(E;\Gamma_t)$ of Eq.~\eqref{eq:rhoGgam} for the four-pole $t\bar b$
channel as a function of $\Gamma_t$. Peak positions are quoted relative
to the stable-limit reference mass $M_{1S}^{(0)}=177.837$~GeV in GeV.
The prominence threshold is $5\%$ of the curve maximum. The count is
operational and depends on the stated prominence criterion. The $t\bar c$
and $t\bar u$ channels exhibit the same progression from multi-peak to
single broad maximum (see text). At small width the four peaks coincide
with the stable pole positions
$\{0,\,0.293,\,0.495,\,0.658\}$~GeV; at the physical width the single
broad maximum sits at $E-M_{1S}^{(0)}\simeq0.382$~GeV, i.e.~absolute
energy $E_{\rm peak}\simeq178.219$~GeV, inside the $t\bar b$ pole band
$[177.837,178.495]$~GeV.}
\label{tab:peak-count}
\begin{tabular}{|c|c|l|}
\hline
$\Gamma_t$ (GeV) & Peak count & Peak positions $E-M_{1S}^{(0)}$ (GeV)\\
\hline
$0.05$ & 4 & $-0.000,\ 0.294,\ 0.496,\ 0.659$\\
$0.10$ & 4 & $0.001,\ 0.294,\ 0.496,\ 0.655$\\
$0.20$ & 3 & $0.010,\ 0.299,\ 0.492$\\
$0.30$ & 1 & $0.319$\\
$0.50$ & 1 & $0.391$\\
$0.70$ & 1 & $0.396$\\
$1.00$ & 1 & $0.391$\\
$1.42$ & 1 & $0.382$\\
\hline
\end{tabular}
\end{table}

\begin{figure}[H]
\centering
\includegraphics[width=0.82\textwidth]{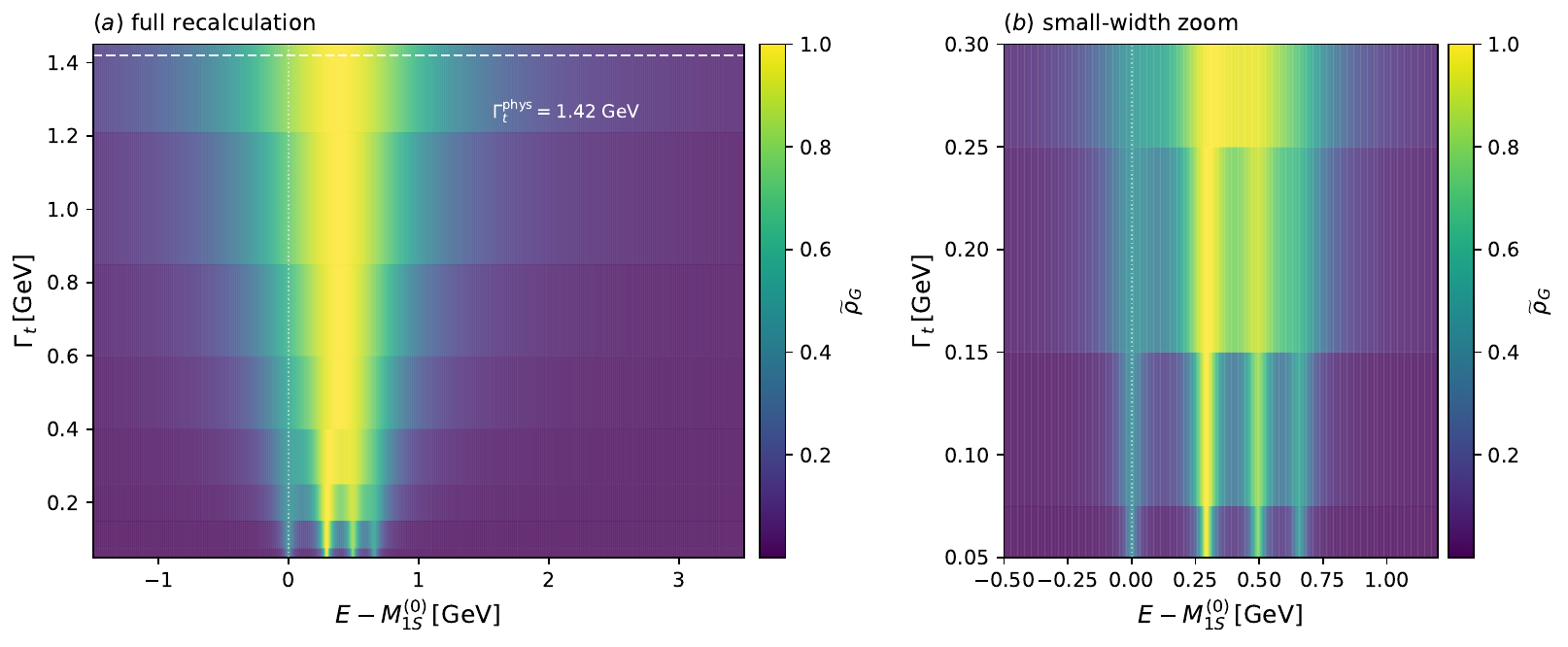}
\caption{
Full width-dependent non-Hermitian evolution of the four lowest
Gaussian-source $t\bar b$ low-pole source-projected responses. At each
$\Gamma_t$ the Salpeter operator $H(m_t-i\Gamma_t/2)$ is
re-diagonalized from scratch, the left and right eigenvectors and the
raw residues $Z_n^{\rm raw}(\Gamma_t)$ are recomputed, and
$\rho_G(E;\Gamma_t)$ is formed from Eq.~\eqref{eq:rhoGgam}. Each curve
is normalized to its own maximum according to Eq.~\eqref{eq:rhotilde}.
Panel (a) shows the full range $0.05\leq\Gamma_t\leq1.42$~GeV; panel
(b) enlarges the small-width region $0.05\leq\Gamma_t\leq0.30$~GeV.
The horizontal axis is $E-M_{1S}^{(0)}$, with $M_{1S}^{(0)}$ the
stable-limit reference mass. The dashed line marks the physical top
width. The separated radial maxima visible at small width progressively
lose resolvability, leaving a single broad maximum near the physical
value.
}
\label{fig:width-evolution}
\end{figure}

\subsection{Source dependence}\label{sec:source-dep}

The single-maximum conclusion at the physical width is unchanged under
variations of the Gaussian source width $\beta$ (scanned at $0.3$,
$0.5$, $0.8$, $1.0$~GeV) and the component-weight ratio $c_2/c_1$
(scanned at $0$, $0.5$, $1$, $2$). The random positive-residue scan
described in Appendix~\ref{app:residue-scan} now plays an auxiliary
role: with the pole-sector source-projected response in hand, the
positive-residue scan is no longer a core piece of evidence but a
robustness check on the normalized positive diagnostic. Detailed
residue tables and peak-count tests are collected in
Appendix~\ref{app:numerics} and Appendix~\ref{app:residue-scan}.

\subsection{Time-domain interpretation}

Resolving two levels separated by $\Delta M$ requires coherence over a
time of order $t_{\rm res}\sim1/\Delta M$. The probability that the top
survives weak decay over that interval is
\begin{equation}
P_{\rm surv}(t_{\rm res})\simeq
\exp(-\Gamma_t t_{\rm res})
=\exp\!\left(-\frac{\Gamma_t}{\Delta M}\right)
\simeq\exp\!\left(-\frac{1}{\mathcal R}\right).
\label{eq:survival}
\end{equation}
For the first radial splitting, Table~\ref{tab:channels} gives a survival
factor of order $10^{-2}$. The definition $t_{\rm res}\sim1/\Delta M$
carries an order-one convention dependence; the individual percentages
are not precision predictions. Equation~\eqref{eq:survival} is a
timescale estimate rather than a top-meson formation probability.

\section{Spin channel and contrast with toponium}\label{sec:contrast}

The independently solved $0^-$ and $1^-$ Salpeter systems give the same
first radial masses and $\mathcal R_1$ at the present $1$~MeV mass
resolution (Appendix~\ref{app:stability}). The full width-dependent
non-Hermitian evolution of Sec.~\ref{sec:dissolution} is carried out on
the $0^-$ channel; the $1^-$ stable-spacing hierarchy has been
independently verified and differs from $0^-$ only at the
$1$~MeV level, so a full $1^-$ width scan is not required unless the
$0^-/1^-$ difference becomes dynamically relevant. The contrast with
toponium is dynamical: for a single-top system with fixed spectator
mass, the radial scale is governed by spectator and QCD dynamics and
does not increase proportionally to $m_t$. For a Coulombic heavy-heavy
system, the characteristic energy scales as $m_t\alpha_s^2$ and is
numerically comparable to the top width. This behavior is consistent
with the recent complex-energy $T$-matrix analysis of
toponium~\cite{Tang:2026zhq} and with precision threshold
calculations~\cite{Nason:2026oka}. The present single-top result does
not contradict the observed $t\bar t$ enhancement; it demonstrates that
the same top width overwhelms the smaller radial scales of the mesonic
reference systems considered here.

\section{Implications for LHC phenomenology}\label{sec:implications}

The present calculation is not a prediction of a measured invariant-mass
distribution. It provides a pole-sector benchmark for process-level
studies in which top-flavoured mesonic correlations are introduced as
effective or diagnostic objects. In such applications, the stable-top
Salpeter eigenvalues should be treated as reference positions of the QCD
correlation problem, rather than as narrow resonances that can be mapped
one-to-one onto LHC peaks.

The relevant partonic topology after the top constituent weak decay is
$t\bar q\to Wb\bar q$. The leading color flow couples the $b$ (which
inherits the top color) with the spectator $\bar q$, so the natural
heavy-flavor final states are:
\begin{center}
\begin{tabular}{|c|c|c|}
\hline
Theory correlation & Final-state heavy-flavor pair & Candidate hadronic final states\\
\hline
$t\bar b$ & $b\bar b$ & bottomonium, open-bottom hadrons\\
$t\bar c$ & $b\bar c$ & $B_c$-like, open heavy-flavor hadrons\\
$t\bar u$ & $b\bar u$ & $B$-mesons, open-bottom hadrons\\
\hline
\end{tabular}
\end{center}
These assignments are qualitative flavor-flow statements, not
predictions of hadronization fractions or exclusive branching ratios.
The production of charmonium from $t\bar c$, for example, would require
additional $c\bar c$ pair creation through gluon splitting, which is not
a direct consequence of the $t\bar c$ correlation studied here.

A complete simulation may include short-distance production coefficients,
nonresonant amplitudes, top decay, parton showering, hadronization, and
detector effects. These ingredients may distort a broad threshold-region
distribution, but they do not alter the bound-state hierarchy identified
here: the inherited top width is much larger than the radial splittings
generated by spectator and QCD dynamics.

The practical implication is therefore a consistency requirement. Any
LHC-oriented calculation that attributes multiple narrow structures
to the stable-top radial eigenvalues should identify the additional mechanism that
overcomes $\Gamma_t\gg\Delta M$ and demonstrate that the effect persists in
a gauge-complete production-and-decay treatment. Concretely, within the isolated single-top pole sector studied here,
any structure narrower than approximately $0.3$~GeV in the $t\bar b$,
$t\bar c$, or $t\bar u$ correlation at the physical top width cannot be
generated by resolving the stable-top radial reference levels alone and
must arise from an additional mechanism---such as short-distance
production dynamics or continuum coupling---capable of overcoming the
$\Gamma_t\gg\Delta M$ hierarchy.

\section{Conclusions}\label{sec:conclusion}

Within the heavy-quark expansion and the constant complex-pole-mass
continuation, an unstable heavy-light system has a parametrically simple
structure: the pole width is inherited from the unstable heavy
constituent up to $\mathcal O(\Lambda_{\rm kin}^2/m_Q^2)$ corrections, while
radial splittings remain $\mathcal O(\Lambda_{\rm rad})$:
\[
\Gamma_n=\Gamma_Q\left[1+O\!\left(\frac{\Lambda_{\rm kin}^2}{m_Q^2}\right)\right],
\qquad
\Delta M_n=\mathcal O(\Lambda_{\rm rad}).
\]
An artificial heavy-mass scaling test confirms the predicted $m_Q^{-2}$
suppression with fitted exponents $p_n\simeq1.95$--$1.99$ and
$R^2>0.999$ for all four radial states.

The top quark, with $\Gamma_t\simeq1.42$~GeV and
$\Lambda_{\rm rad}\sim0.2$--$0.3$~GeV, is the extreme realization of
this hierarchy: $\Gamma_t\gg\Lambda_{\rm rad}$. The full
non-Hermitian Salpeter calculation provides an operator-level
quantitative test: the biorthogonal Hellmann--Feynman relation
$\Gamma_n=\Gamma_t\,\partial M_n/\partial m_t$ is verified at the
$10^{-5}$ level (Table~\ref{tab:hf-verify}), and
$\Gamma_n/\Gamma_t=0.99995$--$1.00000$ for all radial states and all
three flavor channels.

The radial overlap measures lie between $0.11$ and $0.23$, well below
the exact $1/\sqrt{3}$ double-maximum threshold. The low-pole
source-projected response $\rho_G^{(N)}(E)$, constructed from raw
biorthogonal residues without positive-normalizing or real-part
projection, exhibits a single broad maximum at the physical top width in
all three channels ($t\bar b$, $t\bar c$, $t\bar u$), with strictly
positive values (minimum-to-maximum ratio above $+0.04$).

The full width-dependent non-Hermitian re-diagonalization confirms that
this conclusion does not depend on a frozen-residue approximation. At
each $\Gamma_t\in\{0,0.05,\ldots,1.42\}$~GeV the Salpeter operator is
reconstructed from scratch; the separated radial maxima at small width
progressively dissolve into a single broad response near the physical
value, with real pole positions moving by less than $1.4\times10^{-7}$~GeV.
Pole-truncation stability ($N_{\rm pole}=4,6,8$) and a direct
full-matrix resolvent evaluation ($\delta_{\rm shape}=8\times10^{-4}$,
peak agreement $<10^{-5}$~GeV) confirm that no artifact is introduced
by the pole truncation.

The stable-top eigenvalues survive as reference poles, but their radial
organization does not survive as a spectrally resolvable single-top
mesonic spectrum at the physical top width.
Experimentally, the corresponding weak-decay topology would be $Wb\bar q$, but the present
calculation does not predict an exclusive decay channel or a measured
invariant-mass distribution. Its implication is more limited and more
robust: any process-level interpretation of multiple narrow structures
in such final states must invoke dynamics beyond the stable-top radial
pole organization. The Gaussian source is a
controlled resolution probe of the bound-state subspace rather than a
collider production operator. A complete observable prediction requires
short-distance production coefficients, continuum and nonresonant
amplitudes, a gauge-consistent treatment of top decay, parton showering,
hadronization, and detector response. The present complex-pole,
low-pole response, heavy-mass scaling, and full-width-evolution analysis
supplies a bound-state benchmark and a consistency criterion for
process-level calculations that seek to relate observable structures to
stable-top radial reference levels.

\vspace{0.1cm}
{\bf Acknowledgments}

This work was supported by the National Natural Science
Foundation of China under Grants 12575106 and 12147214, and
the Specific Fund of Fundamental Scientific Research Operating
Expenses for Undergraduate Universities in Liaoning Province
under Grant No.~LJ212410165019. The authors acknowledge the
use of ChatGPT and Aether for language and presentation
assistance. All scientific content, calculations, and
conclusions are the authors' responsibility.

\appendix

\section{Numerical robustness}\label{app:numerics}

\begin{table}[htbp]
\centering
\caption{Full $t\bar b$ widths compared with the kinetic estimate.}
\label{tab:widths}
\begin{tabular}{|c|c|c|c|}
\hline
State & $\Gamma_n^{\rm full}$ (GeV) & $\Gamma_n^{\rm kin}$ (GeV)
& $(\Gamma_n^{\rm full}-\Gamma_n^{\rm kin})/\Gamma_t$\\
\hline
$1S$ & 1.41997 & 1.41999 & $-1.2\times10^{-5}$\\
$2S$ & 1.41995 & 1.41999 & $-2.8\times10^{-5}$\\
$3S$ & 1.41994 & 1.42000 & $-3.8\times10^{-5}$\\
$4S$ & 1.41993 & 1.42000 & $-4.6\times10^{-5}$\\
\hline
\end{tabular}
\end{table}

One-parameter-at-a-time variations give
\begin{center}
\begin{tabular}{|c|c|c|c|}
\hline
Parameter & Baseline & Scan range & $\mathcal R_1$ range\\
\hline
$\alpha_s$ & 0.11 & 0.09--0.13 & 0.197--0.217\\
$\lambda$ (GeV$^2$) & 0.18 & 0.12--0.24 & 0.164--0.245\\
$\alpha$ (GeV) & 0.06 & 0.04--0.08 & 0.202--0.211\\
\hline
\end{tabular}
\end{center}
The PDG interval $\Gamma_t=1.27$--$1.61$~GeV gives
$\mathcal R_1=0.182$--$0.231$. Varying $\Delta q$ from 0.006 to
0.016~GeV at $q_{\max}=4$~GeV, or varying $q_{\max}$ from 3 to 5~GeV
at $\Delta q=0.008$~GeV, leaves $M_{1S}=177.8366$~GeV,
$\Delta M=0.2934$~GeV, and $\mathcal R_1=0.2066$ at the quoted precision.

A direct numerical verification of the Hellmann--Feynman relation
(Eq.~\eqref{eq:widthinherit}) is reported in Table~\ref{tab:hf-verify}
in the main text. The symmetric finite-difference step
$\delta m$ was varied over $10^{-4}$--$10^{-2}$~GeV around the stable
point; the mass derivative $\partial M_n/\partial m_t$ is stable to
better than $10^{-5}$ in the quoted digits throughout this interval,
while the relative finite-difference error of the reconstructed width
stays below $10^{-5}$ in units of $\Gamma_t$.
Varying the relative finite-difference step in the heavy-mass scaling
test from $5\times10^{-5}$ to $2\times10^{-4}$ changes the fitted
exponents by less than $0.01$, leaving the $m_Q^{-2}$ scaling
unchanged.

The source-width dependence of the normalized residues is tested by
varying the Gaussian width $\beta$ from $0.3$ to $1.0$~GeV
(Table~\ref{tab:beta-scan}):

\begin{table}[htbp]
\centering
\caption{Normalized Gaussian-source Salpeter residues $\widehat Z_n$
[Eq.~\eqref{eq:Znorm}] for the $t\bar b$ channel at several source
widths $\beta$, with $\Gamma_t=1.42$~GeV and $c_2/c_1=1$.}
\label{tab:beta-scan}
\begin{tabular}{|c|c|c|c|c|}
\hline
$\beta$ (GeV) & $\widehat Z_{1S}$ & $\widehat Z_{2S}$ & $\widehat Z_{3S}$ & $\widehat Z_{4S}$\\
\hline
$0.3$ & 0.080 & 0.252 & 0.322 & 0.346\\
$0.5$ & 0.159 & 0.370 & 0.289 & 0.182\\
$0.8$ & 0.405 & 0.493 & 0.087 & 0.015\\
$1.0$ & 0.610 & 0.420 & $-0.046$ & 0.016\\
\hline
\end{tabular}
\end{table}
The residue pattern changes significantly with $\beta$. For
$\beta\lesssim0.8$~GeV all four normalized residues are positive,
whereas at $\beta=1.0$~GeV the $3S$ residue is negative,
$\widehat Z_{3S}=-0.046$, so the positive pole response of
Eq.~\eqref{eq:rho} no longer admits a strictly positive-weight
interpretation at that source width; the raw low-pole response
$\rho_G^{(N)}(E)$ remains well defined. The observed
redistribution of radial residues reflects the interplay between the
source momentum profile and the nodal structure of the Salpeter wave
functions. Despite this variation, the single-maximum response persists
for all $\beta$ values at the physical width, confirming that the loss
of radial spectroscopy is not an artifact of a particular source-width
choice.

The component-weight ratio $c_2/c_1$ in Eq.~\eqref{eq:source} controls the
relative coupling of the Gaussian envelope to the two independent $0^-$
Salpeter radial amplitudes $f_1(q)$ and $f_2(q)$. It is scanned at
$c_2/c_1=0$, $0.5$, $1$, and $2$ with $\beta=0.5$~GeV fixed. The
normalized residues redistribute appreciably across this range, but the
physical-width response retains a single broad maximum in all four cases,
as verified by automatic peak counting from sign changes of
$d\rho^{(+)}/dE$ with a prominence threshold. This confirms that the
single-maximum conclusion is not tied to the symmetric component choice
$c_2/c_1=1$ used for the main figures.

\section{Positive-residue scan}\label{app:residue-scan}

\begin{figure}[H]
\centering
\includegraphics[width=0.82\textwidth]{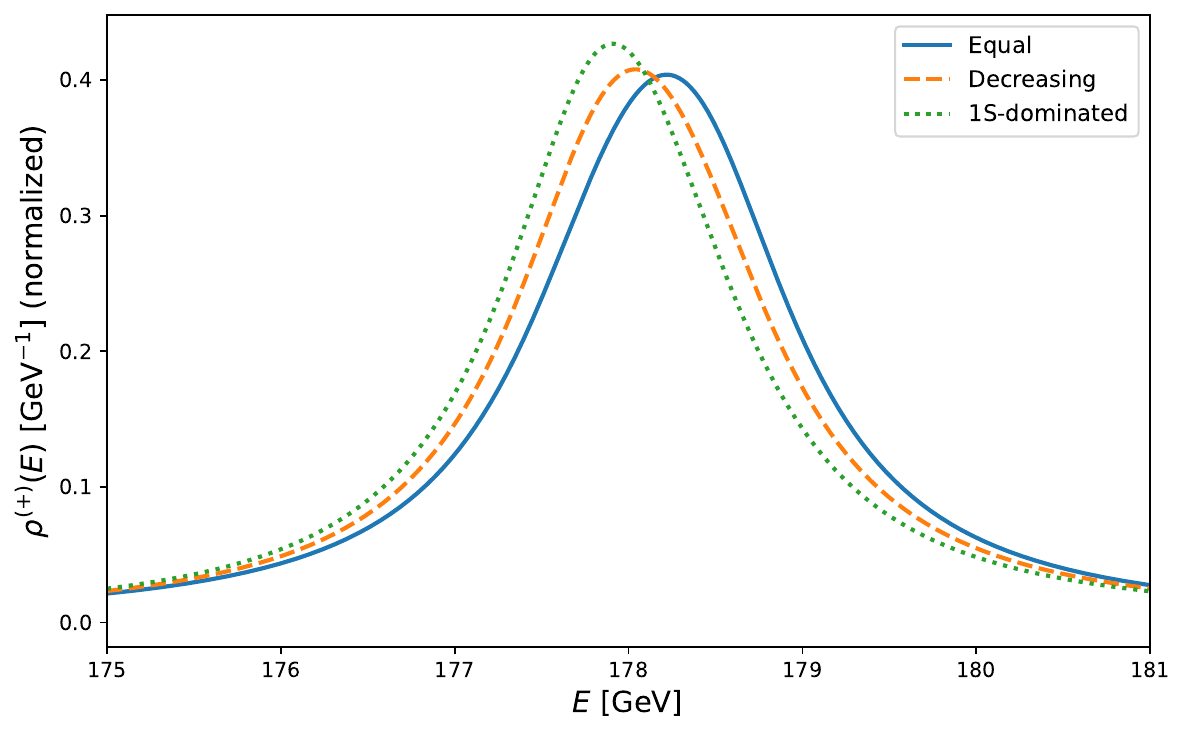}
\caption{Benchmark positive pole responses for three representative
residue patterns, each normalized to unit area. These are not computed
from Salpeter eigenvectors; they serve as a robustness check
complementing the Gaussian-source results in Sec.~\ref{sec:response}.}
\label{fig:app-residue}
\end{figure}

The residue robustness test uses the four $t\bar b$ masses and widths
in Table~\ref{tab:tbpoles}. Each trial begins with independent weights
$w_n$ drawn log-uniformly from $10^{-12}\leq w_n\leq1$, followed by the
normalization $Z_n=w_n/\sum_k w_k$. The scan was implemented in Python
using NumPy's random generator with seed $42$, and
\texttt{scipy.signal.find\_peaks} for peak counting. The response in
Eq.~\eqref{eq:rho}
is evaluated on $4001$ equally spaced energy points. All $2\times10^5$
sampled quadruplets contain exactly one interior maximum. Figure~\ref{fig:app-residue}
shows three representative benchmark patterns for illustration. This
numerical test supports the robustness of the positive-response conclusion
but does not constrain line-shape distortions generated by complex
amplitudes or nonresonant backgrounds.

\section{Spin channel and non-normal eigenvalue stability}
\label{app:stability}

The $0^-$ and $1^-$ equations are implemented separately with the same
kernel and complex-mass replacement. At the present $1$~MeV output
resolution, the $1S$ masses and $\mathcal R_1$ values coincide:
\begin{center}
\begin{tabular}{|c|c|c|c|c|c|}
\hline
Channel & $M_{1S}^{0^-}$ & $M_{1S}^{1^-}$ & $|\Delta_{\rm HF}|$ & $\mathcal R_1^{0^-}$ & $\mathcal R_1^{1^-}$\\
& (GeV) & (GeV) & (MeV) & & \\
\hline
$t\bar b$ & 177.837 & 177.837 & $<1$ & 0.207 & 0.207\\
$t\bar c$ & 174.662 & 174.662 & $<1$ & 0.230 & 0.230\\
$t\bar u$ & 173.437 & 173.437 & $<1$ & 0.221 & 0.221\\
\hline
\end{tabular}
\end{center}

For right and left eigenvectors $HR_n=\Mcal_nR_n$,
$H^\dagger L_n=\Mcal_n^*L_n$, $L_n^\dagger R_n=1$, we use
$\kappa_n=\|L_n\|_2\|R_n\|_2$. For the $t\bar b$ $0^-$ states at the
physical width,
\begin{center}
\begin{tabular}{|c|c|c|c|c|}
\hline
State & $\kappa_n(0)$ & $\kappa_n(\Gamma_t)$ & $r_n^R$ & $r_n^L$\\
\hline
$1S$ & 1.99789 & 1.99789 & $3\times10^{-15}$ & $7\times10^{-15}$\\
$2S$ & 3.30808 & 3.30808 & $3\times10^{-15}$ & $8\times10^{-15}$\\
$3S$ & 4.05329 & 4.05329 & $7\times10^{-15}$ & $1\times10^{-14}$\\
$4S$ & 4.76598 & 4.76598 & $9\times10^{-15}$ & $2\times10^{-14}$\\
\hline
\end{tabular}
\end{center}
The relative changes in $\kappa_n$ between zero and physical width are
below $3\times10^{-8}$, expected because the leading width contribution is
approximately $-i\Gamma_t\mathbb I/2$. The small departure from the
identity, of order $\Lambda_{\rm kin}^2/m_t^2$, establishes the
heavy-top approximation as a quantitatively valid description at a
level far below the radial splitting scale; the full
re-diagonalization quantifies this validity explicitly rather than
relying on an \emph{a priori} smallness argument.

The near-degeneracy of the $0^-$ and $1^-$ masses at the $1$~MeV
level is consistent with the expectation that the hyperfine splitting,
which scales as $\langle q^2\rangle/m_t^2$ in the heavy-top limit, is
strongly suppressed relative to the radial splittings and is further
subdominant to the top width. The screened Cornell potential's
spin-spin contact term contributes at a scale below the current output
precision; the coincidence of the two channels confirms that the
hyperfine structure is dynamically negligible for the radial
resolvability question studied here.

The peak-counting procedure used in Sec.~\ref{sec:dissolution} employs
\texttt{scipy.signal.find\_peaks} with a relative prominence threshold
of $5\%$ of each curve's maximum value and a minimum separation of five
grid points (corresponding to $\Delta E\simeq0.006$~GeV). These
parameters were chosen so that a resolved peak is defined as a local
maximum whose prominence is at least $5\%$ of the dominant peak, which
is insensitive to grid-scale fluctuations while preserving genuine
multi-peak structure at small $\Gamma_t$. Varying the relative prominence
threshold from $3\%$ to $10\%$ does not alter the transition from a
multi-peak pattern at small width to a single maximum at the physical
width.

The channel metadata and pole inputs used in the positive-response
calculations were checked explicitly against
Table~\ref{tab:tbpoles}: the four $0^-$ $t\bar b$ poles lie at
$M_{1S}=177.837$, $M_{2S}=178.130$, $M_{3S}=178.332$, and
$M_{4S}=178.495$~GeV. The low-pole response in
Fig.~\ref{fig:pole-response} is constructed from these poles using the
full complex Gaussian-source residues defined in Eq.~\eqref{eq:source-residue}.
Their real parts are listed in Table~\ref{tab:residues-raw}.

\section{Full width-dependent evolution and pole truncation}
\label{app:full-width}

This appendix collects the per-state width evolution
(Table~\ref{tab:widths-app}), the full-vs-frozen-residue comparison at
representative widths, the state-tracking diagnostics, and the
pole-truncation stability test.

\begin{table}[htbp]
\centering
\caption{Full width-dependent non-Hermitian re-diagonalization for the
four lowest $0^-$ $t\bar b$ radial states at representative widths.
$\operatorname{Re}\Mcal_n(\Gamma_t)$ and
$\Gamma_n(\Gamma_t)=-2\operatorname{Im}\Mcal_n(\Gamma_t)$ are obtained
by re-solving the biorthogonal eigenproblem at each $\Gamma_t$; the
ratio $\Gamma_n/\Gamma_t$ is the direct test of the width-inheritance
relation $\Gamma_n\simeq\Gamma_t$ of Eq.~\eqref{eq:widthinherit}.}
\label{tab:widths-app}
\begin{tabular}{|c|c|c|c|c|}
\hline
$\Gamma_t$ (GeV) & state & $\operatorname{Re}\Mcal_n$ (GeV) &
$\Gamma_n$ (GeV) & $\Gamma_n/\Gamma_t$\\
\hline
$0.10$ & $1S$ & $177.8366$ & $0.1000$ & $1.0000$\\
$0.10$ & $2S$ & $178.1299$ & $0.1000$ & $1.0000$\\
$0.10$ & $3S$ & $178.3317$ & $0.1000$ & $1.0000$\\
$0.10$ & $4S$ & $178.4948$ & $0.1000$ & $1.0000$\\
\hline
$0.20$ & $1S$ & $177.8366$ & $0.2000$ & $1.0000$\\
$0.20$ & $2S$ & $178.1299$ & $0.2000$ & $1.0000$\\
$0.20$ & $3S$ & $178.3317$ & $0.2000$ & $1.0000$\\
$0.20$ & $4S$ & $178.4948$ & $0.2000$ & $1.0000$\\
\hline
$0.50$ & $1S$ & $177.8366$ & $0.5000$ & $1.0000$\\
$0.50$ & $2S$ & $178.1299$ & $0.5000$ & $1.0000$\\
$0.50$ & $3S$ & $178.3317$ & $0.5000$ & $1.0000$\\
$0.50$ & $4S$ & $178.4948$ & $0.5000$ & $1.0000$\\
\hline
$1.42$ & $1S$ & $177.8366$ & $1.4200$ & $0.99998$\\
$1.42$ & $2S$ & $178.1299$ & $1.4200$ & $0.99997$\\
$1.42$ & $3S$ & $178.3317$ & $1.4199$ & $0.99996$\\
$1.42$ & $4S$ & $178.4948$ & $1.4199$ & $0.99995$\\
\hline
\end{tabular}
\end{table}

The maximum displacement $|\operatorname{Re}\Mcal_n(\Gamma_t^{\rm phys})-M_n^{(0)}|$
across the entire scan is $1.4\times10^{-7}$~GeV for $t\bar b$,
$7.5\times10^{-8}$~GeV for $t\bar c$, and $3.4\times10^{-8}$~GeV for
$t\bar u$. The real pole positions are essentially unchanged, while the
widths inherit the top width to one part in $10^4$. State tracking via
the biorthogonal overlap
$\mathcal O_{mn}=|L_m^\dagger(\Gamma_i)R_n(\Gamma_{i+1})|$ gives the
identity assignment across all eight transitions in all three channels,
with maximum off-diagonal overlap
$4\times10^{-5}$ ($t\bar b$), $2\times10^{-5}$ ($t\bar c$), and
$9\times10^{-6}$ ($t\bar u$). No state scrambling occurs.

The frozen-residue approximation of Eq.~\eqref{eq:rhofrozen} captures
the qualitative dissolution pattern (multi-peak at small width, single
broad maximum at the physical width) but shifts the peak position
slightly relative to the full recalculation. The two prescriptions agree
that the physical-width response has a single broad maximum; we therefore
report only the full-recalculation result in the main text and retain the
frozen-residue scan as an auxiliary quantitative cross-check.

\subsection*{Direct full-matrix resolvent check}

\begin{figure}[H]
\centering
\includegraphics[width=0.82\textwidth]{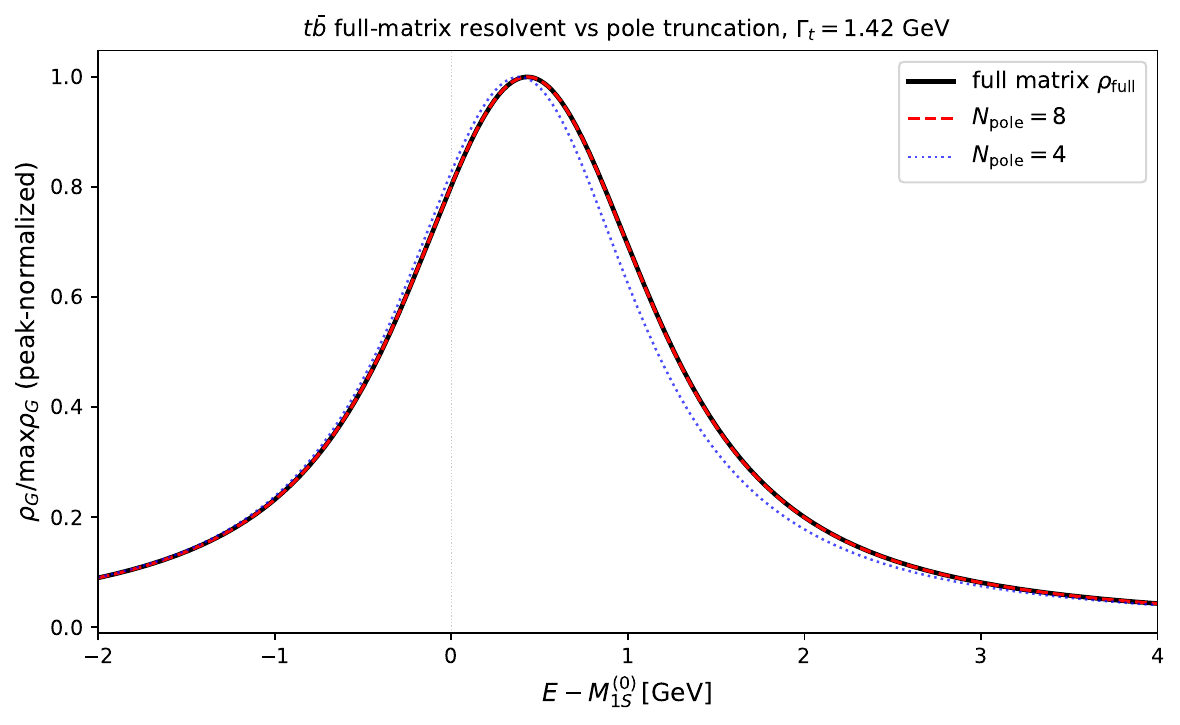}
\caption{Direct full-matrix resolvent $\rho_{\rm full}(E)$ (solid black)
compared with the $N_{\rm pole}=8$ (dashed red) and $N_{\rm pole}=4$
(dotted blue) pole-truncated representations, for the $t\bar b$ channel
at $\Gamma_t=1.42$~GeV. All three curves are peak-normalized for shape
comparison; the horizontal axis is $E-M_{1S}^{(0)}\,[{\rm GeV}]$. The
full matrix and the eight-pole representation are visually
indistinguishable on this scale.}
\label{fig:full-vs-N8}
\end{figure}

The main analysis evaluates $\rho_G^{(N)}(E)$ via the truncated pole sum
of Eq.~\eqref{eq:rhoG}. Since the full discretized Salpeter matrix $H$
is available, we also compute the resolvent directly by solving the
linear system $(EI-H)x=s$ at each energy $E$ on the same grid, giving
$G_{\rm full}(E)=s^\dagger x$ and
$\rho_{\rm full}(E)=-(1/\pi)\operatorname{Im}G_{\rm full}(E)$. This is
a resolvent of the finite discretized Salpeter space only; it does not
include physical continuum contributions, which would require a
process-level correlation function as noted in Sec.~\ref{sec:resolvent-recon}.
We use \texttt{scipy.linalg.solve} for the general complex matrix; the
residual $\|(EI-H)x-s\|_2/\|s\|_2$ stays below $2\times10^{-13}$ across
the entire grid. Figure~\ref{fig:full-vs-N8} compares $\rho_{\rm full}$
with the $N_{\rm pole}=8$ and $N_{\rm pole}=4$ representations. A direct
evaluation of $s^\dagger(E-H)^{-1}s$ with the full discretized Salpeter
matrix confirms the single-maximum physical-width response and agrees
with the eight-pole representation within
$\delta_{\rm shape}=\max_E|\rho_{\rm full}/\max\rho_{\rm full}
-\rho_8/\max\rho_8|=8\times10^{-4}$ over the displayed energy window;
the main peak positions coincide to within $10^{-5}$~GeV. The four-pole
representation already captures the peak structure but deviates by
$\sim7\%$ in the tails, reflecting the $\sim5\%$ residue weight carried
by the $5S$--$8S$ modes.

\begin{figure}[H]
\centering
\includegraphics[width=0.82\textwidth]{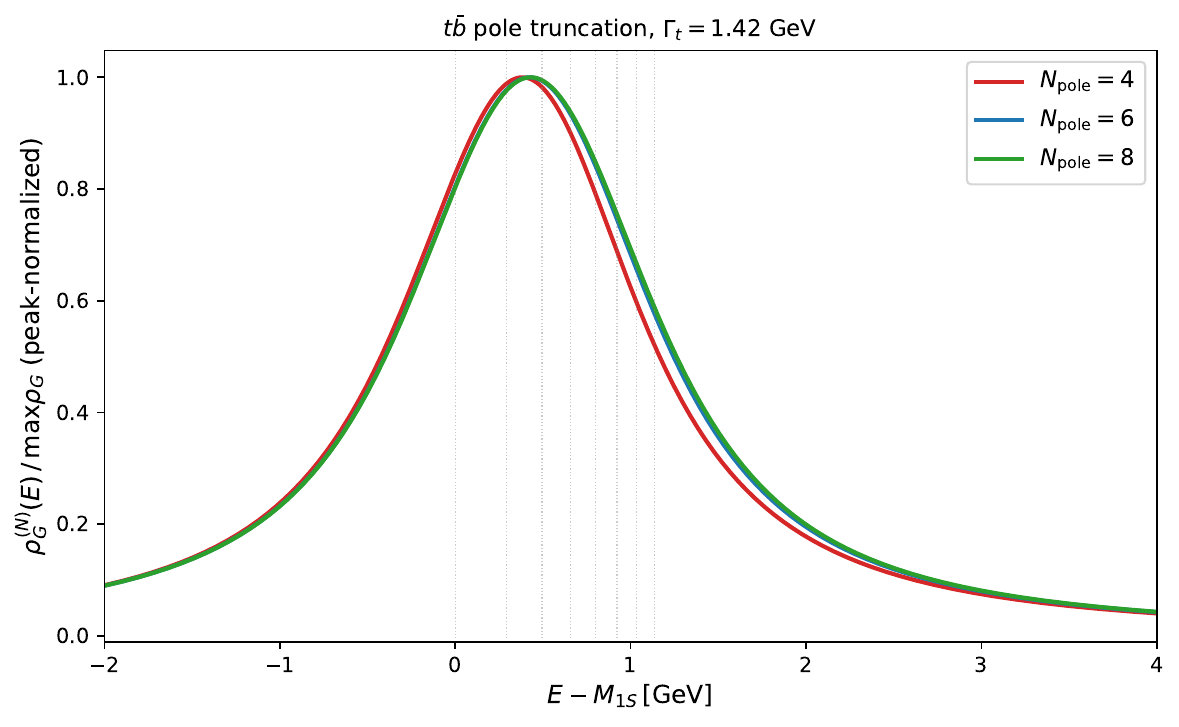}
\caption{Pole-truncation stability test for the $t\bar b$ channel at
the physical width. The low-pole response
$\rho_G^{(N)}(E)$ is evaluated with $N=4$, $6$, and $8$ physical poles
and peak-normalized for shape comparison. Adding $5S$--$8S$ (which carry
a combined $\sim5\%$ additional residue weight) does not introduce any
new resolvable radial peak. The main peak position shifts by
$0.048$~GeV ($3.4\%$ of $\Gamma_t$) between $N=4$ and $N=8$, far below
the radial splitting scale.}
\label{fig:pole-truncation}
\end{figure}

Figure~\ref{fig:pole-truncation} shows the pole-truncation stability
test. All three truncations ($N=4$, $6$, $8$) give a single broad
maximum in the same energy region; the main peak position shifts by
$0.048$~GeV between $N=4$ and $N=8$, i.e.~$3.4\%$ of $\Gamma_t$ and far
below the radial splitting scale of $0.2$--$0.3$~GeV. Higher poles
mostly affect the tails and the overall normalization; they do not
restore a resolvable radial multiplet. The biorthogonality and
eigen-residual diagnostics remain at the $10^{-14}$ level for the
eight-pole basis.

\bibliographystyle{unsrt}
\bibliography{references}
\end{document}